\documentclass[fleqn,usenatbib]{mnras}

\usepackage{newtxtext,newtxmath}
\usepackage[T1]{fontenc}

\DeclareRobustCommand{\VAN}[3]{#2}
\let\VANthebibliography\thebibliography
\def\thebibliography{\DeclareRobustCommand{\VAN}[3]{##3}\VANthebibliography}

\usepackage{physics}
\usepackage{color}
\usepackage{graphicx}
\graphicspath{{./images/}}
\usepackage{afterpage}
\usepackage{color}
\usepackage{hyperref}
\usepackage{blindtext}
\usepackage{multicol}
\newcommand{\marvel}{\textsc{MARVEL}}
\newcommand{\pgopher}{\textsc{Pgopher}}
\newcommand{\exocross}{\textsc{ExoCross}}
\newcommand{\duo}{\textsc{Duo}}
\newcommand{\molpro}{\textsc{Molpro}}
\newcommand{\hitran}{\textsc{Hitran}}
\newcommand{\LineList}{\textsc{SOLIS}}

\newcommand{\ai}{\textit{ab initio}}
\newcommand{\Ai}{\textit{Ab initio}}

\newcommand{\Xstate}{\ensuremath{X\,{}^{3}\Sigma^{-}}}
\newcommand{\astate}{\ensuremath{a\,{}^{1}\Delta}}
\newcommand{\bstate}{\ensuremath{b\,{}^{1}\Sigma^{+}}}
\newcommand{\cstate}{\ensuremath{c\,{}^{1}\Sigma^{-}}}
\newcommand{\dstate}{\ensuremath{d\,{}^{1}\Pi}}
\newcommand{\estate}{\ensuremath{e\,{}^{1}\Pi}}
\newcommand{\Astate}{\ensuremath{A\,{}^{3}\Pi}}
\newcommand{\Bstate}{\ensuremath{B\,{}^{3}\Sigma^{-}}}
\newcommand{\Cstate}{\ensuremath{C\,{}^{3}\Pi}}
\newcommand{\Cprimestate}{\ensuremath{C'\,{}^{3}\Pi}}
\newcommand{\Aprimestate}{\ensuremath{A'\,{}^{3}\Delta}}
\newcommand{\Aprimeprimestate}{\ensuremath{A''\,{}^{3}\Sigma^{+}}}

\newcommand{\tsub}[1]{\textsubscript{#1}} %shorter subscript
\newcommand{\wav}[1]{#1 cm$^{-1}$}
\newcommand{\cm}{cm$^{-1}$}

\newcommand{\band}[2]{$#1\xrightarrow[]{}#2$}
\newcommand{\brkt}[3]{$\bra{#1}#2\ket{#3}$}
\newcommand{\brkteq}[3]{\bra{#1}#2\ket{#3}}

\newcommand{\SO}{$^{32}$S$^{16}$O}

\title[ExoMol line lists -- {LVI}. SO]{ExoMol line lists -- {LVI}:  The SO line list, MARVEL analysis of experimental transition data and  refinement of the  spectroscopic model}

\author[Brady et al.]{
Ryan P. Brady,$^{1}$
Sergei N. Yurchenko,$^{1}$
Jonathan Tennyson,$^{1}$\thanks{The corresponding author: j.tennyson@ucl.ac.uk}
Gap-Sue Kim,$^{2}$
\\
$^1$ Department of Physics and Astronomy, University College London, Gower Street, WC1E 6BT London, UK \\
$^2$ Dharma College, Dongguk University, 30, Pildong-ro 1-gil, Jung-gu, Seoul 04620, Korea \\
}

\date{Accepted XXXX. Received XXXX; in original form XXXX}

\date{\today}

\begin{document}

\label{firstpage}

\maketitle

\pagerange{\pageref{firstpage}--\pageref{lastpage}}

\begin{abstract}
A semi-empirical IR/Vis line list, SOLIS, for the sulphur monoxide molecule  $^{32}$S$^{16}$O is presented. SOLIS includes accurate empirical rovibrational energy levels, uncertainties, lifetimes, quantum number assignments, and transition probabilities in the form of Einstein $A$ coefficients 
covering the \Xstate$, \astate, \bstate, \Astate, \Bstate, \Aprimeprimestate, \Aprimestate$\ and \estate\ systems and wavenumber range up to 43303.5~cm$^{-1}$ ($\geq 230.93$ nm) with $J\le 69$. SOLIS has been computed by solving the rovibronic Schr\"{o}dinger equation for diatomics using the general purpose variational code \textsc{Duo} and starting from a published \textit{ab initio} spectroscopic model of SO (including potential energy curves, coupling curves, (transition) dipole moment curves) which is refined to experimental data. To this end, a database of 50~106 experimental transitions, 48~972 being non-redundant, has been compiled through the analysis of 29 experimental sources, 
and a self-consistent network of 8558 rovibronic energy levels for the $X$, $a$, $b$, $A$, $B$, and $C$ electronic states has been generated with the \textsc{MARVEL} algorithm covering rotational and vibrational quantum numbers $J \leq 69$ and $v \leq 30$ and energies up to 52350.40~cm$^{-1}$. No observed transitions connect to the \Bstate$(v = 0)$ state which is required to model perturbations correctly, so we leave fitting the $B\,{}^3\Sigma^-$ and $C\,{}^3\Pi$ state UV model to a future project. The SO line list is available at ExoMol from \href{www.exomol.com}{www.exomol.com}. 

\end{abstract}


\section{Introduction}
\label{sec:Introduction}

Sulphur monoxide ($^{32}$S$^{16}$O) is an often  transient diatomic molecule whose spectral properties are important for a wide range of environments. SO was first detected by radio astronomy in the interstellar medium by means of rotational spectroscopy and was the first $^3\Sigma$ ground state molecule to be detected in outer space \citep{73GoBaxx.SO}. Additionally, SO was the first molecule observed using pure rotational transitions using microwave spectroscopy within its excited electronic states \citep{70Saito.SO}. Since its first detection, SO has been observed in many astronomical environments, including the interstellar medium \citep{78GoGoLi.SO}, molecular clouds \citep{97CoMuxx.SO,87BlSuMa.ISM}, and planetary and lunar atmospheres \citep{96Lellouch, 02deRoGr.SO, 90NaEsSk.SO, 12BeMoBe.SO}. Numerous studies propose sulphur-bearing molecules, including SO, as constituents of volcanic planetary atmospheres \citep{98ZoFexx.SO,21HoRiSh, 12Krasnopolsky}. It also plays a role in many solar-system atmospheres, including that of Jupiter’s moon Io \citep{96Lellouch, 02deRoGr.SO} and of Venus \citep{90NaEsSk.SO, 12BeMoBe.SO}, as well as Earth's own atmosphere \citep{87BuLoHa.SO}. The recent detection of SO$_2$ in at atmosphere of WASP-39b \citep{23RuSiMu} suggested  SO as part of the photochemical production paths of SO$_2$ \citep{23TsKePo.SO2}. SO also has importance for: (1) experimental applications in spectroscopy, such as UV lasing \citep{91MiYaSm.SO, 92StCaPo.SO}; (2) astrophysical applications, such as shock modelling \citep{96Bachiller.ISM, 93ChMaxx.SO, 91AmElEl.SO}, magnetic field measurements in molecular clouds through  observation of SO Zeemen splitting \citep{74ClJoxx.SO, 17CaLaCo.SO}, and studies into planetary formation mechanisms after its first detection within  a protoplanetary disk \citep{15PaFuAg}, which supports sulphur-carrier based dust-formation channels  \citep{19LaJaCa.S,17ViLoJa.SO}; (3) telluric applications, such as environmental studies of acid rain, since it is an intermediate in combustion reactions and has great chemical involvement with with N\tsub{2} and O\tsub{2} \citep{48Gaydon.book,87BuLoHa.SO}.

The rovibrational structure of SO’s spectrum for the \Xstate\, \astate, \bstate, \Astate, and \Bstate\ electronic states at low vibrational excitation's has been studied by numerous works; we provide a full analysis of the experimental coverage on SO below. SO's electronic transitions were first reported by \citet{32Martin.SO}, and has since been subject to pure rotational \citep{94KlBeSa.SO,93Yamamoto.SO}, electronic \citep{68Colin.SO,82Colin.SO, 99SeFiRa.SO}, and ro-vibrational \citep{85KaBuKa.SO, 82WoAmBe.SO, 88KaTiHi.SO} spectroscopic studies. The pure rotational transitions within several vibrational states in the ground \Xstate\ electronic state have been measured in the terahertz \citep{15MaHiMo.SO, 97KlBeWi.SO}, far infrared \citep{76ClDexx.SO, 94CaClCo.SO}, and microwave regions \citep{93Yamamoto.SO, 92LoSuOg.SO, 97BoCiDe.SO}.  For most of these spectra, SO was studied in non-local thermodynamic equilibrium (non-LTE) conditions, and so only relative intensities are available at best. Currently, there are no  published absolute intensity measurements for SO. However, measurements of state lifetimes provide information on Einstein A coefficients and hence transition dipole moments \citep{16TeHuNa.method}. Experimental lifetimes for the \bstate, \Astate\ and \Bstate\ states have been measured \citep{99ElWexx.SO, 69Smith.SO, 03YaTaTo.SO}, and provide a valuable benchmark for our intensity calculations (see Section \ref{subsec:int_scale}).

In this study we present the largest compilation of experimental transition data and derived self-consistent empirical rovibrational energy levels for \SO\ to date. The derived energy levels where obtained through use of the \marvel\ (Measured-Active-Rotational-Vibrational-Energy-Levels) spectroscopic network algorithm, to which we format the \SO\ data based on the \marvel\ format \citep{MARVEL}. We then refine our \ai\ spectroscopic model \citep{22BrYuKi.SO} to our determined empirical energies to produce a hot semi-empirical line list \LineList\ for \SO\  as part of the ExoMol project \citep{12TeYuxx.exo, 20TeYuAl}. The \LineList\ line list supplements existing spectroscopic line list data for SO which are limited in coverage. For example, spectroscopic databases CDMS \citep{CDMS} and NIST \citep{NIST}  databases contain data covering the microwave region only. HITRAN \citep{jt857}  considers relatively low vibrational excitations for transitions between electronic states \Xstate, \astate\ and \bstate\ only. We compare the \LineList\ line list to the existing spectra data in Section \ref{subsec:comp_to_exp_spec}.

\section{Theoretical Background}
\label{Sec:Theory_Background}
\subsection{The \marvel\ Procedure}
\label{subsec:MARVEL}

The critical evaluation of experimental transition data and formation of a self-consistent set of rovibronic energy levels is done through the \marvel\ procedure \citep{MARVEL,12FuCsi.method} which is built on the concept of spectroscopic networks (SN) \citep{12FuCsxx.methods,11CsFuxx.methods}. Through a weighted linear least squares protocol, \marvel\ inverts the information contained within transition data to form a set of associated energy levels and uncertainties. Self consistency within the energy levels is then achieved through an iterative re-weighting algorithm which adjusts (increases) the uncertainties in the line positions to an optimised uncertainty $\sigma_{\rm opt}$ until they agree with the rest of the network.

The inverted \marvel\ energy levels form nodes of a SN, which are linked by transitions, to which the validation of experimental information can be done on all data simultaneously using elements of network theory. The final energy level uncertainties in the SN are obtained through combining the optimal \marvel\ uncertainties of all transitions connecting a given energy level. This study used a new implementation, \marvel\ 4, which uses a bootstrap method to determine uncertainties in the empirical energy levels it determines \citep{jt908}. We used 100 iterations with the bootstrap method to determine the uncertainties in our empirical MARVEL energy levels.

\subsection{Quantum Numbers}
\label{subsec:QN_selection_rules}

We assign the rovibronic energy levels of \SO\ using the vibrational and rotational quantum numbers $v$ and $J$,  respectively, rotationless parity  $\tau$ ($e/f$) and, in line with Hund's case-(a) coupling scheme and  sublevels denoted by the fine structure $F_{2S+1}$. For $J\geq \Lambda+S$, the fine structure, $F_{2S+1}$, is defined for triplet electronic states via
\begin{equation}
\begin{split}
F_1 & = N+S, \\
F_2 & = N, \\
F_3 & = N-S,
\end{split}
\end{equation}
where singlet states have no spin projection (i.e. $F_2$) and the total angular momentum excluding electronic and nuclear spin is labelled $N$. For linear molecules such as diatomics, levels with parity $+(-1)^{J}$ and $-(-1)^{J}$ are labelled $e$ and $f$ levels, respectively, and their relation to the $\pm$ parities are given in Table \ref{tab:ef_pm}. We thus assign every experimental rovibrational transition using the  $v,J,e/f,F$ quantum numbers and standard spectroscopic notation for electronic states. 
For nuclear motion calculations we use the quantum numbers $\Lambda$, $\Sigma$, and $\Omega$ to assign electronic states, which are the projection of orbital, spin, and total angular momentum on the bond axis, respectively, in additional to the state labels \Xstate, \Astate, \bstate\ etc. For the \Xstate\ and \Bstate\ states the spin-parity sub-levels in increasing energy order are
\begin{equation}
\begin{split}
(F_1,e) &: \Lambda =0, \; \Sigma =0,\; \Omega =0, \\
(F_2,f) &: \Lambda =0, \; \Sigma =1,\; \Omega =1, \\
(F_3,e) &: \Lambda =0, \; \Sigma =1,\; \Omega =1,
\end{split}
\end{equation}
whereas for the regular \Astate\ state there is lambda-doubling and we have
\begin{equation}
\begin{split}
(F_1,e/f) &: \Lambda =1, \; \Sigma =-1,\; \Omega =0, \\
(F_2,e/f) &: \Lambda =1, \; \Sigma =\hphantom{+}0,\; \Omega =1, \\
(F_3,e/f) &: \Lambda =1, \; \Sigma =+1,\; \Omega =2
\end{split}
\end{equation}
and for the inverted \Cstate\ state the sublevels increase in energy with decreasing $\Omega$
\begin{equation}
\begin{split}
(F_1,e/f) &: \Lambda =1, \; \Sigma =+1,\; \Omega =2, \\
(F_2,e/f) &: \Lambda =1, \; \Sigma =\hphantom{+}0,\; \Omega =1, \\
(F_3,e/f) &: \Lambda =1, \; \Sigma =-1,\; \Omega =0.
\end{split}
\end{equation}
Rigorous electric-dipole selection rules hold here, and can be summarised as
\begin{align*}
          + \leftrightarrow\ -, & \\
        \Delta J = \pm 1 & (e\leftrightarrow\, e, f\xrightarrow{}f),\\
                    \Delta J = 0 & (e\leftrightarrow\,f,  0 \not\leftrightarrow\, 0 )
\end{align*}

\begin{table}
\footnotesize
    \centering
    \caption{Relation between the $e/f$ and $\pm$ parities for linear molecules.}
    \label{tab:ef_pm}
    \begin{tabular}{ccc}
        \hline
        $e/f$ & $J$ & $\pm$ \\ 
        \hline\hline 
        $e$ & even & $+$ \\
        $e$ & odd & $-$ \\
        $f$ & even & $-$ \\
        $f$ & odd & $+$ \\
        \hline
    \end{tabular}
\end{table}

\section{The Experimental Transition Database}
\label{sec:exp_database}

\subsection{Outline}
\label{subsec:outline}
Table \ref{tab:exp_database} summarises the experimental transition data included within our \marvel\ analysis where each study is conveniently labelled with a tag including the first two digits of the year of publication and letters of the names of the first three authors in the form 'YYAaBbCc'. Table \ref{tab:exp_database} includes the spectral coverage of each study, the associated quantum number coverage of their assignments, and the mean uncertainty of their results. We compiled a total of 50~106 transitions, of which 49~613 are non-redundant, from 29 experimental sources covering the \Xstate, \astate, \bstate, \Astate, \Bstate\, and \Cstate\ electronic states of SO for rovibrational excitation's $J\leq 69$, $v \leq 30$.

\subsection{General Comments}
\label{subsec:general_comments}

A crucial limitation of the experimental data set for SO (Table \ref{tab:exp_database}) is in the vibrational state coverage of the lower electronic states. Transitions to/within states beyond the third vibrational excitation for  \Xstate, \astate, \bstate, and \Astate\ are severely lacking. Some vibrational transition data involving states beyond $v > 3$ are available, but inclusion of these within our \marvel\ analysis often led to fragmented SNs. 

Our literature review found that no  transitions have been measured  associated with the  vibrational ground state  of the  \Bstate\ state, which makes it difficult to constrain its PEC minima during refinement of the spectroscopic model. The impact of this missing data is then amplified since the \Bstate\ and \Cstate\ energies exhibit many mutual perturbations  because of their overlapping potentials and strong coupling. To correctly model the perturbations one requires accurate positioning of the potentials corresponding to the resonating states relative to each other, which is made difficult because of the absence of data connecting to the \Bstate$(v=0)$ state.

Isotopologues of SO have been experimentally measured by several sources covering  $^{33}$S$^{16}$O \citep{96KlSaBe.SO, 15MaHiMo.SO,22HeStLy.SO}, $^{34}$S$^{16}$O \citep{97KlBeWi.SO,74Tiemann.SO,82BoDeDe.SO,93Yamamoto.SO,15MaHiMo.SO,87BuLoHa.SO,82Tiemann.SO,22HeStLy.SO}, $^{32}$S$^{17}$O \citep{96KlSaBe.SO}, $^{32}$S$^{18}$O \citep{96KlSaBe.SO,97KlBeWi.SO,74Tiemann.SO,82BoDeDe.SO,87BuLoHa.SO,82Tiemann.SO} and the rare isotopologue $^{36}$S$^{16}$O \citet{22HeStLy.SO,96KlSaBe.SO}. Few studies measure transitions within excited electronic states for these isotopologues, where \citet{97KlBeWi.SO} measured the \astate\ and \bstate\ states, \citet{93Yamamoto.SO} measured the \bstate\ state, and \citet{22HeStLy.SO} measured the higher \Astate, \Bstate, and \Cstate\ electronic states. Low vibronic excitation is typically measured with similar $J$ coverage as the main \SO\ isotopologue. 

\subsection{Source Specific Comments}
\label{subsec:source_specific_comments}

\begin{enumerate}

\renewcommand{\theenumi}{(\alph{enumi})}

\item  A significant problem we faced during data analysis is that several literature sources did not provide obvious uncertainties on their line measurements, which is important for their validation within the \marvel\ protocol (Section \ref{subsec:MARVEL}). We thus had to estimate their uncertainties through combination difference (CD) tests to other data in our database with known uncertainties where possible. These sources include \citet{71BoMaxx.SO, 72BoMaxx.SO, 69Colin.SO, 82Colin.SO} and \citet{94StCaPo.SO}. Initial uncertainties for these sources were assumed to be \wav{0.05} and manually increased with successive \marvel\ runs until the data gave satisfactory CD relations  with other sources. In the case of blended lines, their uncertainties were increased by a factor of $\sqrt{2}$ relative to the non-blended data which often resulted in their validation. As a result of this the source uncertainties were estimated to be \wav{0.02} \citep{94StCaPo.SO}, \wav{0.4} \citep{71BoMaJa.SO, 72BoMaJa.SO}, \wav{0.05} \citep{69Colin.SO} and  \wav{0.2} \citep{82Colin.SO}. \label{l:a}

\item Another issue with the experimental transition data is the significant proportion of blended lines that are reported, such  in the measurements by \citet{69Colin.SO, 82Colin.SO, 86ClTexx.SO,87BuLoHa.SO,88KaTiHi.SO} and \citet{97BoCiDe.SO}. To account for potential inaccuracy in their assignments the blended lines were given a lower weight in our SN model (see comment (a) above). \label{l:b}

\item The experimental sources \citet{82WuMoYe.SO, 85KaBuKa.SO, 87EnKaHi.SO, 88KaTiHi.SO, 93Yamamoto.SO, 94StCaPo.SO, 96KlSaBe.SO, 97BoCiDe.SO, 97KlBeWi.SO} and \citet{03KiYaxx.SO} provide transition data in high vibrational states which were removed from the MARVEL SN (see Section \ref{subsec:general_comments}). \label{l:c}

\item 15MaHiMo \citep{15MaHiMo.SO} provide much data on SO isotopologues with determined isotopically invariant parameters as well as other various constants for the lowest 7 vibrational states. \label{l:d}

\item 87BuLoHa \citep{87BuLoHa.SO} contains a misprint in Table 1 column 5, the SO (\Xstate, $v=1-0$, $R(18)$) line should be \wav{1108.81665}, not \wav{1008.81665}. \label{l:e}

\item 88KaTiHi \citep{88KaTiHi.SO} provide 60 SO (\astate, $v=3-4, 4-5$) transitions. If high vibrational data for SO becomes available these would be a prime source for inter-vibrational transition data to supplement our \marvel\ dataset. \label{l:f}

\item A discrepancy in the CDMS data for the \Xstate, $v=1$ state of SO was found. The lowest $v=1$ state energy had to be shifted by  \wav{26.4559}. Furthermore, a shift of \wav{6.478} was found in the \astate, $v=0$ CDMS data of SO, where the source of error may come from use of a high uncertainty  $v=0-0$ band centre due to \citet{84BiElFi}. The relative energies between CDMS levels within the same vibrational levels are unaffected, hence the transition wavenumbers are correct, but were corrected before being used in analyses involving \marvel. \label{l:g}

\item \citet{22HeStLy.SO} perform high-resolution FUV Fourier-transform photoabsorption spectroscopy and provide the only published UV transition data covering the \Cstate$\leftarrow$\Xstate and \Bstate\ 
 ($v=4\ldots 30$) $\leftarrow$ \Xstate\ bands. Because of the large overlap and spin-orbit coupling (SOC) between  the \Bstate\ and \Cstate\ states, many perturbations are present within the experimental data which appear to be assigned accurately. They also provide transition data for the isotopologues $^{33}$S$^{16}$O and $^{36}$S$^{16}$O. \label{l:h}

 \item \citet{99SeFiRa.SO} provide 74 \bstate\ $\--$ \Xstate\ magnetic dipole transitions, which have the same selection rules as for electric dipole transitions except from the parity changing rule. We do not include these in our MARVEL network. \label{l:i}
  
\item We chose to omit the 540 \Astate\;$(v^{\prime}=2)\--$\Xstate\;$(v^{\prime\prime}=0)$ transitions measured by \citet{69Colin.SO} for two reasons: (1) they produced many conflicts with the more comprehensive and more accurate data by \citet{22HeStLy.SO}; (2) use of  MARVELised energies generated including these data for refining our spectroscopic model proved very difficult;  we prescribed abnormalities in the energy structure to be due to the poor data which did not occur when using the equivalent MARVELised data from \citet{22HeStLy.SO}. \label{l:j}

\end{enumerate}

\begin{figure}
    \centering
    \includegraphics[width=\linewidth]{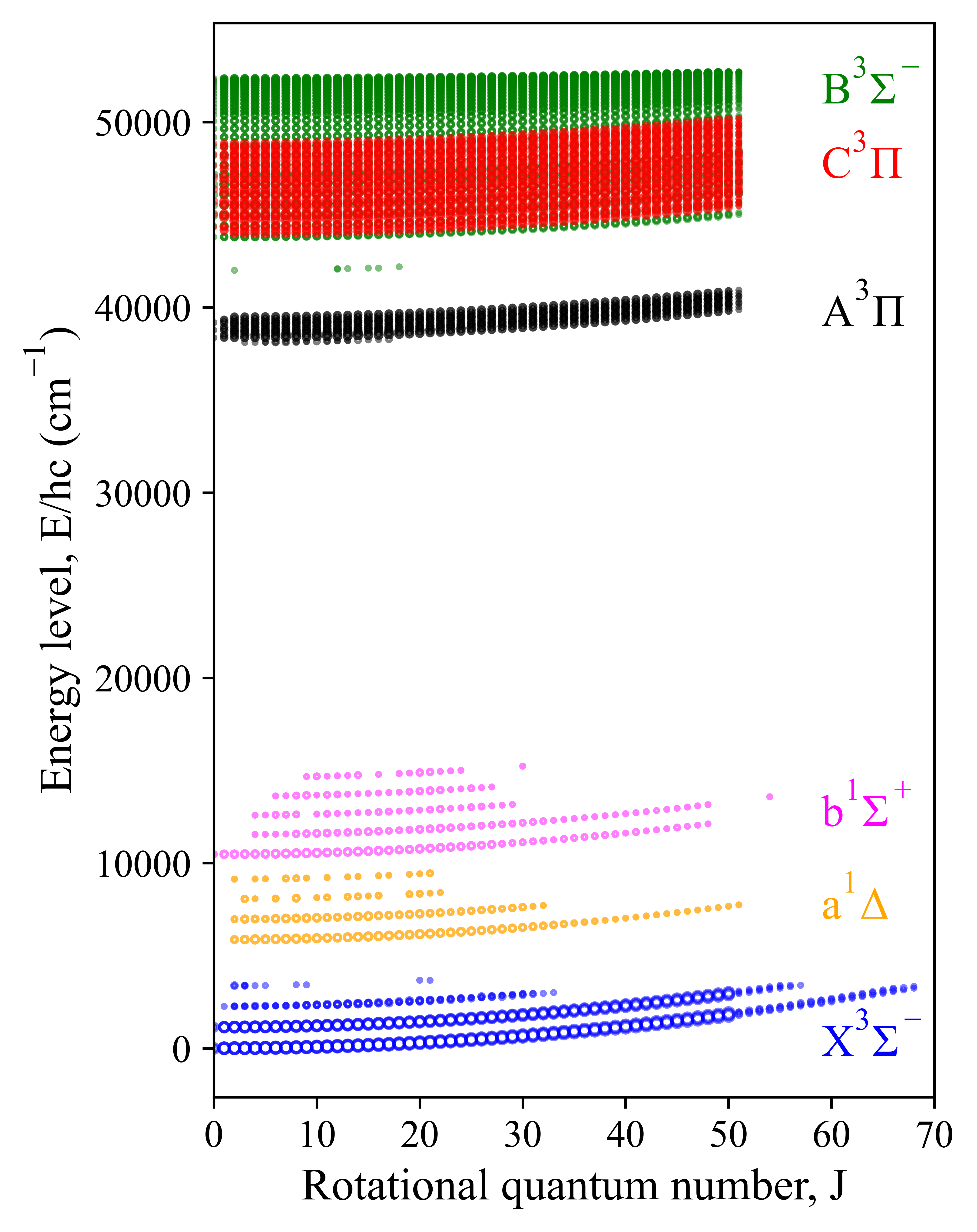}
    \caption{Our generated MARVEL energies plotted against the angular momentum quantum number $J$ for the \Xstate, \astate, \bstate, \Astate, \Bstate, and \Cstate\ states. The vertical structure within each electronic state corresponds to the different vibrational levels. The size of the plot markers are directly proportional to the number of combination differences to that level in the SN.}
    \label{fig:MARVEL_eners}
\end{figure}

\begin{table*}
\def\arraystretch{0.8}
\footnotesize
    \centering
    \caption{The experimental data sources included in the final MARVEL analysis and their spectroscopic coverage. TAG denotes the identifier used to label the data sources throughout this paper, V/T describes the number of validated or included (V) data using the \marvel\ procedure described in Section \ref{subsec:MARVEL} relative to the total number of provided transitions (T), and the final column cross-references source specific comments (Comm) in Section \ref{subsec:source_specific_comments}.}
    \label{tab:exp_database}
    \begin{tabular}{llllllllr}
        \hline
        \multicolumn{1}{l}{TAG} &  \multicolumn{1}{c}{Source} & \multicolumn{1}{l}{Range (cm$^{-1}$)} & \multicolumn{1}{c}{El. states} & \multicolumn{1}{l}{$J$} & \multicolumn{1}{l}{$v$}  &  \multicolumn{1}{l}{ $\bar{\sigma}$ (cm$^{-1}$)} & \multicolumn{1}{l}{V/T} & \multicolumn{1}{r}{Comm} \\
        \hline\hline
 64PoLi & \citet{64PoLixx.SO} & 0.435-2.2 & $X$-$X$ & 0-3 & 0-0                   & 1.07$\times 10^{-5}$ & 5/5 &   \\
 64WiGoSa &\citet{64WiGoSa.SO} & 2.87-5.74 & $X$-$X$ & 1-4 & 0-0                   & 1.67$\times 10^{-5}$ & 6/6 &   \\
 69Colin &\citet{69Colin.SO} & 38672.94-39086.99 & $A$-$X$ & 0-34 & 0-2            & 0.06 & 0/514 &  \ref{l:a},\ref{l:b},\ref{l:j} \\
 71BoMa &\citet{71BoMaxx.SO} & 11354.43-11606.78 & $b$-$X$ & 5-96 & 0-2          & 0.40 & 154/227 &  \ref{l:a} \\
 72BoMa &\citet{72BoMaxx.SO} & 12265.5-12625.29 & $b$-$X$ & 8-32 & 0-4           & 0.40 & 123/165 &  \ref{l:a} \\
 74Tiemann &\citet{74Tiemann.SO} & 1.21-4.31 & $X$-$X$ & 1-4 & 0-0                 & 8.64$\times 10^{-8}$ & 6/6 &   \\
 76ClDe &\citet{76ClDexx.SO} & 4.26-11.6 & $X$-$X$, $a$-$a$ & 0-9 & 0-0            & 4.50$\times 10^{-5}$ & 28/28 &   \\
 82Colin &\citet{82Colin.SO} & 38255.26-39499.71 & $A$-$X$ & 1-34 & 0-1            & 0.21 & 246/275 &  \ref{l:a},\ref{l:b} \\
 82WoAmBe &\citet{82WoAmBe.SO} & 3368.19-3386.30 & $X$-$X$ & 1-9 & 0-3         & 2.00$\times 10^{-3}$ & 28/28 &   \\
 82WuMoYe &\citet{82WuMoYe.SO} & 23696.68-40816.33 & $B$-$X$ & 0-0 & 1-19    & 2.00$\times 10^{-4}$ & 0/9 & \ref{l:c} \\
 82Tiemann &\citet{82Tiemann.SO} & 1.21-9.89 & $X$-$X$ & 1-9 & 0-0                 & 3.53$\times 10^{-5}$ & 5/5 &   \\
 85KaBuKa &\citet{85KaBuKa.SO} & 1041.95-1116.20 & $X$-$X$ & 1-44 & 0-6        & 2.14$\times 10^{-3}$ & 50/94 & \ref{l:c} \\
 86ClTe &\citet{86ClTexx.SO} & 38051.24-38108.07 & $A$-$X$ & 1-24 & 0-0          & 0.15 & 74/87 & \ref{l:b}  \\
 87BuLoHa &\citet{87BuLoHa.SO} & 1051.89-2296.98 & $X$-$X$, $a$-$a$ & 0-47 & 0-2 & 1.64$\times 10^{-3}$ & 560/562 & \ref{l:b},\ref{l:e}  \\
 87EnKaHi &\citet{87EnKaHi.SO} & 10.9-12.8 & $a$-$a$ & 7-9 & 0-5                   & 9.00$\times 10^{-7}$ & 12/24 & \ref{l:c} \\
 88KaTiHi &\citet{88KaTiHi.SO} & 1022.14-1121.26 & $a$-$a$ & 2-41 & 0-5        & 2.01$\times 10^{-3}$ & 82/144 & \ref{l:b},\ref{l:c},\ref{l:f}  \\
 92LoSuOg &\citet{92LoSuOg.SO} & 0.435-0.435 & $X$-$X$ & 1-1 & 0-0                 & 6.67$\times 10^{-7}$ & 0/1 &   \\
 93Yamamoto &\citet{93Yamamoto.SO} & 2.8-15.4 & $b$-$b$ & 1-11 & 0-8               & 5.89$\times 10^{-7}$ & 42/42 & \ref{l:c}  \\
 94CaClCo &\citet{94CaClCo.SO} & 19-62.8 & $X$-$X$, $a$-$a$ & 9-45 & 0-0               & 1.46$\times 10^{-4}$ & 33/33 &   \\
 94StCaPo &\citet{94StCaPo.SO} & 39619.44-40280.32 & $A$-$X$, $B$-$X$ & 1-26 & 0-5.    & 0.03 & 85/237 &  \ref{l:a},\ref{l:c} \\
 96KlSaBe &\citet{96KlSaBe.SO} & 19.7-34.4 & $X$-$X$ & 12-25 & 0-7                 & 3.76$\times 10^{-6}$ & 45/71 &  \ref{l:c} \\
 97BoCiDe &\citet{97BoCiDe.SO} & 11.7-31.2 & $a$-$a$, $b$-$b$ & 8-22 & 0-13            & 1.40$\times 10^{-6}$ & 81/143 & \ref{l:b},\ref{l:c}  \\
 97KlBeWi &\citet{97KlBeWi.SO} & 9.94-35.4 & $a$-$a$, $b$-$b$ & 6-25 & 0-7             & 4.43$\times 10^{-6}$ & 41/55 & \ref{l:c} \\
99SeFiRa &\citet{99SeFiRa.SO} & 5792.97-10566.42 & $a$-$X$, $b$-$X$ & 0-50 & 0-2    & 0.01 & 813/890 &\ref{l:i}  \\
03KiYa &\citet{03KiYaxx.SO} & 1.11-2.8 & $b$-$b$ & 0-2 & 0-22                     & 6.67$\times 10^{-8}$ & 30/30 & \ref{l:c} \\
15MaHiMo &\citet{15MaHiMo.SO} & 0.435-83.8 & $X$-$X$ & 0-60 & 0-0                 & 2.08$\times 10^{-6}$ & 110/110 & \ref{l:d}  \\
17CaLaCo &\citet{17CaLaCo.SO} & 2.87-28.1 & $X$-$X$ & 0-20 & 0-0                  & 6.67$\times 10^{-5}$ & 19/19 &   \\
 CDMS &\citet{CDMS} & 0.43-125.40 & $X$-$X$, $a$-$a$ & 0-69 & 0-1             & 3.97$\times10^{-2}$ & 860/862 &  \ref{l:g} \\
 22HeStLy &\citet{22HeStLy.SO} & 37856.62-52350.40 & $A$-$X$, $B$-$X$, $C$-$X$ & 0-51 & 0-30 & 0.05 & 45434/45434 & \ref{l:h} \\
        \hline\hline
    \end{tabular}
\end{table*}

\subsection{MARVELisation of the Experimental Transition Data}
\label{subsec:marvelisation}

The \SO\ spectroscopic network was built through input of 50~106 rovibronic transitions into \marvel\ from the 29 data sources outlined in Table \ref{tab:exp_database}. 546 transitions were invalidated since their optimised uncertainties did not  satisfy the validation condition $\sigma_{\rm opt}-\sigma_{\rm exp}< 0.05$ \wav{}. Invalidation of transition data can be due to multiple reasons, errors in their quantum number assignment, in their measurement, in the digitization of their scanned data Tables (especially in old papers), and simply because they are not self-consistent with the rest of the network. The latter is the most common cause of invalidation but since they usually connect few levels they are invalidated if the aforementioned reasons are not the cause. The invalidated transitions are removed from the \marvel\ network but are kept in the \marvel\ input file with a negative wavenumber transition frequency. We note the 590 transitions were excluded not because of invalidation through the MARVEL procedure and are detailed in comments \ref{l:i} and \ref{l:j} in Section \ref{subsec:source_specific_comments}.

The majority of transitions that were invalidated are for lines connecting $v > 3$ (69\%) because of the lack of inter-vibrational data energetically above $v > 3$ which, if included, resulted in the fragmentation of our central SN and the invalidation of otherwise seemingly reliable data sources. There are a lack of measurements of rotational transitions within these higher vibrational states. It was found that keeping data for $v\leq3$ produced the largest set of self consistent energy levels and hence SN. For $v\leq3$  the experimental source that provided the most invalidated transitions is by \citet{69Colin.SO} (44\%) who measured the only \band{\Astate}{\Xstate} $v = 2 \xrightarrow{} 0$ band transitions, which is important for the refinement of the \Astate\ potential energy curve (PEC). \citet{69Colin.SO} measured \SO\ in emission by means of flash photolysis of sulphur bearing gases using a medium resolution quartz spectrograph. They provide no direct uncertainty on their line positions, but provide an uncertainty for their \band{\Xstate}{\Bstate} bandheads of $\pm$\wav{1} obtained in their absorption study. If one uses this value as a metric for their line position uncertainties, then it is to be expected that data coming from \citet{69Colin.SO} should be treated with caution, consequently leading to much of their data being invalidated. The majority of the remaining invalidated $v \leq 3$ data comes from \citet{82Colin.SO} and \citet{94StCaPo.SO} (21\% and 22\%, respectively) who measure \band{\Xstate}{\Astate} and \band{\Xstate}{\Bstate} transition bands, respectively. 

As a result of the critical evaluation of the experimental transition data, we invert and provide optimised uncertainties for 8558 rovibrational energy levels for \SO\ which forms a fully self-consistent SN. Figure \ref{fig:MARVEL_eners} plots the \marvel\ energies versus the rotational quantum number $J$, where a large gap in the $\sim$\wav{15~000-37~000} region exists, corresponding to missing highly excited vibrational data in the \Xstate, \astate, and \bstate\ states and any experimental coverage of the intermediate electronic states \cstate, \Aprimestate, and \Aprimeprimestate. For the higher vibrational energy levels of each electronic state there are also gaps in the rotational structure. The experimental transition frequencies collected as part of this work are provided in
the Supporting Information to this paper in the MARVEL format alongside the resulting MARVEL empirical energy levels.

\section{The Spectroscopic Model}
\label{sec:The Spectroscopic Model}
To produce the final semi-empirical line list for \SO\ we use the \ai\ spectroscopic model presented in our recent work \citep{22BrYuKi.SO} as a theoretical baseline for refinement to our MARVELised energy levels. Section \ref{subsec:ai_model} overviews the details of the \ai\ model and Section \ref{subsec:refinment} details the method used to refine it.

\subsection{The \Ai\ Spectrosopic Model}
\label{subsec:ai_model}

The \ai\ curves that make up our SO model \citep{22BrYuKi.SO}  were computed using internally contracted multireference configuration interaction (ic-MRCI) level of theory and aug-cc-pV5Z basis sets. The active space and state averaging were chosen to have 12 active electrons with occupied (8,3,3,0) and closed (4,1,1,0) orbitals under C$_{2v}$ symmetry. The model covers 13 electronic states: \Xstate, \astate, \bstate, \cstate, \Aprimeprimestate, \Aprimestate, \Astate, \Bstate, \Cstate, \dstate, \estate, \Cprimestate, and $(3)^{1}\Pi$, which range up to \wav{66800}, vastly beyond the scope of experimental coverage of the molecule. The \ai\ model includes  13 potential energy curves, 29 dipole and transition dipole moment curves, 25 spin-orbit curves, and 18 electronic angular momentum curves on a grid of points over bond lengths 1-3 \AA, where a diabatic representation was built by removing the avoided crossings between the spatially degenerate pairs \Cstate -- \Cprimestate\ and \estate -- $(3)^1\Pi$ through a property-based diabatisation method. 

\subsection{Refinement}
\label{subsec:refinment}

\begin{figure}
    \centering
    \includegraphics[width=\linewidth]{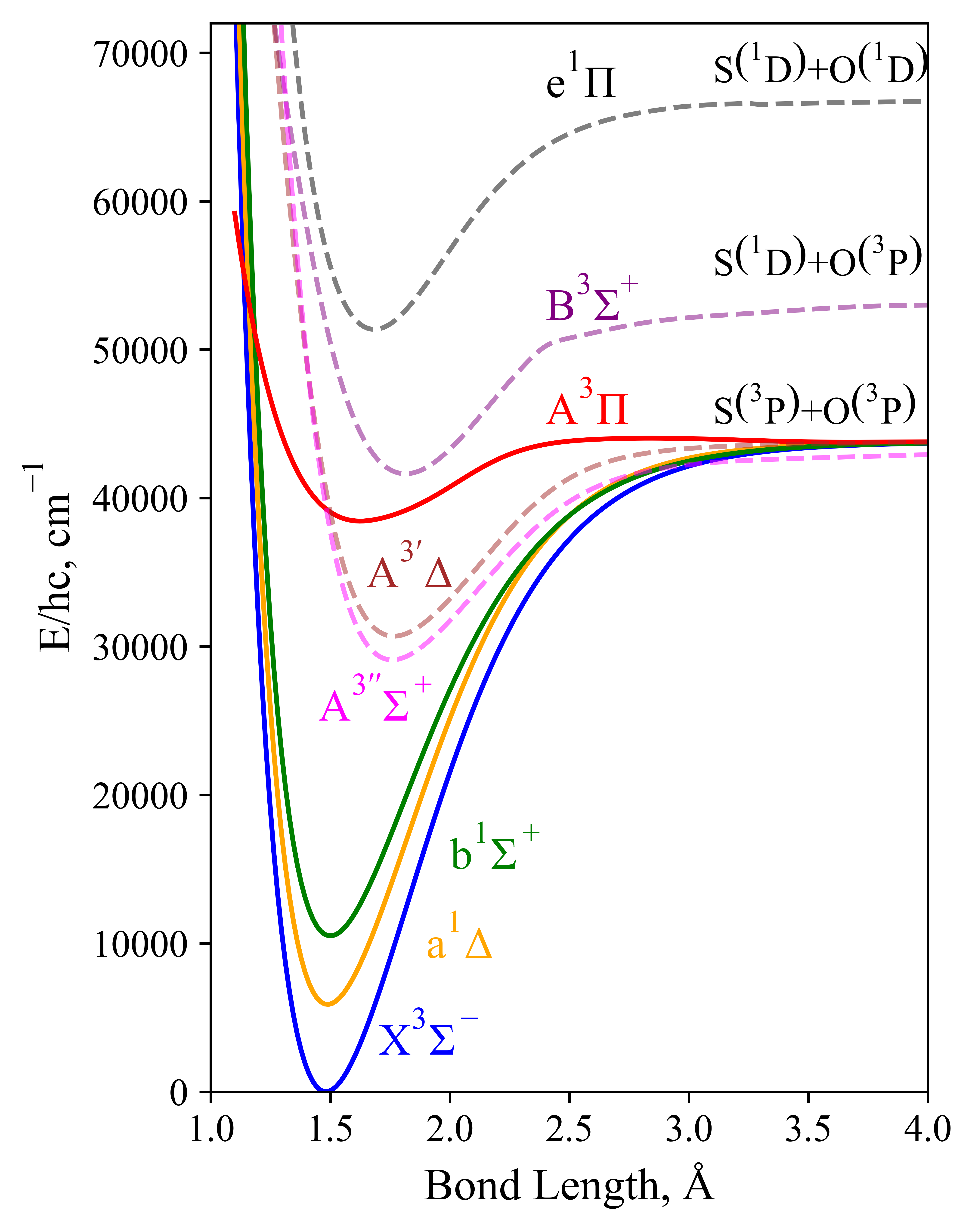}
    \caption{Potential energy curves of states included within our spectroscopic model of SO. The solid lines correspond to the  potentials refined to \marvel\ data, dashed lines correspond to states included within our model but have not been refined, where couplings to these states are essential for the accuracy of the \Xstate, \astate, \bstate, and \Astate\ energies and band intensities.}
    \label{fig:PECs}
\end{figure}

We refined our model to the IR/Vis region by fitting to the \Xstate, \astate, \bstate, and \Astate\ energies only. We include the minimum number of states and couplings required such that our computed energies for these states agree with the \marvel\ energies. Our refined model includes PECs and couplings connecting the \Xstate, \astate, \bstate, \Astate, \Bstate$^\dagger$, \Aprimestate$^\dagger$, \Aprimeprimestate$^\dagger$, \estate$^\dagger$ states, see Fig.~\ref{fig:PECs}, where  potentials for states labelled with a '$\dagger$' superscript are  not refined; these are included solely for their couplings within our model, but their dipoles are kept \ai. These couplings, despite not being the dominant contributions to the energies of the  \Xstate, \astate, \bstate\ and \Astate\ states, will redistribute intensities according to the intensity stealing mechanism (see Sec.\ref{sec:Line List} and the Appendix). Firstly, we found that including the \cstate\ and \dstate\ states  within the model had negligible effect on the energies of the states of interest, and so they were omitted. Secondly, the inclusion of the \Cstate\  state proved problematic and only worsened the fit. Because the \Bstate\ and \Cstate\  PECs have large overlap and strong couplings between each other, their \marvel\ energies include many perturbations because of resonances with each other, and also due to multiple avoided crossings with upper electronic states. Furthermore, due to the lack of important experimental data and consequent lack of proper constraint on the \Bstate\ PEC minimum, efforts  to include the UV region within our fit proved too difficult to do satisfactorily. We therefore removed the \Cstate\ state entirely from our model, but found including the \Bstate, now without resonances with \Cstate, improved the fit of the \Xstate, \astate, \bstate, and \Astate\ energies without being a major contribution to the accuracy of our model. Our initial efforts to fit the UV model constrained the \Bstate\ PEC enough such that expectation values of its couplings to other states in our model were sensible. Furthermore, the current need for the IR/Vis SO line list means we leave work on the UV model to a future study when the appropriate data becomes available.

Refinement of the \ai\ model is facilitated through \duo, a general purpose variational (open access\footnote{\href{https://github.com/Exomol/Duo)}{github.com/Exomol}}) code  that solves the rovibronic Schr\"{o}dinger equation for diatomic molecules. A description of the methodologies used in \duo\ is given by \citet{16YuLoTe.methods}. The refinement process goes as follows: (1) represent PECs, SOCs, EAMCs, DMCs, and other empirically fitted couplings such as rotational Born-Oppenheimer breakdown, spin-spin, and spin-rotational curves with analytical forms; (2) compute energy levels using \duo\ through solving the rovibronic Schr\"{o}dinger equation with curves defined; (3) fit parameters of the analytical functions such that the computed energy levels agree with the \marvel\ energies.

Before refinement, the \marvel\ energy level quantum number assignments need to be converted to the \duo\ quantum numbers $J$, $\tau$, $v$, $\Lambda$, $\Sigma$, and $\Omega$ (see Section \ref{subsec:QN_selection_rules}). Next, the $i$-th \marvel\ energy was given a weight equal to
\begin{equation}
    w_{i} = |\log_{10}(\sigma_i^{\rm opt})|\times n_{i}^{\rm CD}.
\end{equation}
where $\sigma_i^{\rm opt}$ is its optimised uncertainty and $n_{i}^{\rm CD}$ is the number of combination differences/frequency of occurrence within the transition database. This weight is  used in the \duo\ fit meaning energy levels with large uncertainty and a low number of combination differences will have less effect on the fit. Next, a running number must be defined to enumerate the global order of the energy levels calculated by \duo. These should agree with \marvel’s energy ordering for lower $v$ and $J$ numbers, but for higher energy states where \duo’s calculated energies deviate significantly from the \marvel\ data, the ordering of states between the two can differ. The running number we employ increases by 1 per vibrational excitation and by 100 per each electronic state, for example the running numbers for \Xstate($v=1$) and \astate($v=2$) are 1 and 102, respectively. This produced a sensible enumeration that effectively separated the energy levels and provided correct assignments of the calculated energy levels.

\subsubsection{Potential energy, spin–orbit, electronic angular momentum curves}
\label{subsubsec:PEC_SOC_EAMC}
We represent all PECs using the Extended Morse Oscillator (EMO) function \citep{EMO} which has the form
\begin{equation}
\label{eq:EMO}
    V(r) = V_{\rm e} + (A_{\rm e} - V_{\rm e}) \left( 1 - \exp[-\left(\sum_{i=0}^N a_i {\xi_p (r)}^i\right) (r-r_{\rm e}) ] \right)^2,
\end{equation}
where $D_{\rm e}=A_{\rm e}-V_{\rm e}$ is the dissociation energy, $V_{\rm e}$ is the potential minimum and $A_{\rm e}$ is the asymptote, $a_i$ are the expansion coefficients, $r_{\rm e}$  is the equilibrium bond length, and $\xi_p(r)$ is the so called \v{S}urkus variable \citep{84SuRaBo.method} given by
\begin{equation}
    \xi_p(r) = \frac{r^p - r_{\rm e}^p}{r^p + r_{\rm e}^p}
\end{equation}
with $p$ as an integer parameter to allow for a better convergence at large bond lengths values. The \Xstate, \astate, \bstate, \Astate, \Aprimestate, and \Aprimeprimestate\ states dissociate to the same asymptote S(\textsuperscript{3}P) + O(\textsuperscript{3}P), which we initially set at 5.429 eV as reported by \citet{79HeHuxx.book} and then floated during our fits, which converged to a nearby value of 5.42895 eV. The \Bstate\ state adiabatically correlates to S(\textsuperscript{1}D) + O(\textsuperscript{3}P) which we set to a value of 6.5731 eV (\wav{53015.86}) as determined through atomic energies from the NIST atomic database. We shifted the PECs to the \Xstate\ minimum such that $V_{\rm e} (\Xstate)=$ \wav{0}. Since only the \Xstate, \astate, \bstate, \Astate, \Bstate\, and \Cstate\ have experimental transition data to refine their PECs to, the \ai\ PECs for the \Aprimeprimestate, and \Aprimestate\ were fitted to the EMO function given in Eq.~\eqref{eq:EMO} using 10 expansion parameters which ensured accurate representation of their shape as given by \ai\ calculations. Once we fitted the EMO functions, we could then tune their dissociation assymptotes to the 5.429 eV limit \citep{79HeHuxx.book}. The $T_{\rm e}$ value for the  \Aprimeprimestate\ state was fixed to \wav{30~692} as provided by \citet{89Norwood.SO}. We chose not to tune the \estate\  \ai\ PEC  because it has a strong influence on the computed \Xstate, \astate, \bstate, and \Astate\ energies and negatively effects our refinement when altering its potential. This negative effect is due to: (1) The shape of the PEC would be destroyed in tuning the $T_{\rm e}$ and $D_{\rm e}$ values; (2) the \estate\ state has been diabatised \citep{22BrYuKi.SO}, so tuning its PEC would change the avoided crossing morphology and hence a new diabatisation of the spectroscopic model would be required; without experimental data covering \estate\ we chose to keep the diabatised \ai\ potential values.

In our refinement  the \ai\ \Bstate\ PEC  by \citet{19SaNaxx.SO} was used instead of the PEC from our recent \ai\ work \citep{22BrYuKi.SO}. The latter  did not employ sulphur specific diffuse functions, which led to the underestimation of the S(\textsuperscript{3}P) + O(\textsuperscript{3}P), S(\textsuperscript{1}D) + O(\textsuperscript{3}P) and S(\textsuperscript{1}D) + O(\textsuperscript{1}D) dissociation asymptotes and effected the \Bstate\ PEC the worst. The adiabatic character of the \Bstate\ PEC was modelled by diagonalising  a $2\times 2$ matrix of diabatic potentials and the corresponding diabatic coupling which gives the adiabatic potential through solution of the quadratic equation for this system \citep{22BrYuKi.SO}. 

The \ai\ SOCs and EAMCs were morphed from the grid representation to a \v{S}urkus-like expansion \citep{17PrJaLo, 17YuSiLo} given by
\begin{equation}\label{eq:polynom_decay_24}
    F(r)=\sum_{i=0}^{N}B_{i}z^{i}(r)(r-\xi_p)+\xi_{p}B_{\infty},
\end{equation}
where $B_i$ are the expansion coefficients, $B_{\infty}$ is usually taken as zero in-order to allow the expansion to not diverge towards $r\xrightarrow{}\infty$, and $z$ is a damped displacement coordinate  given by
\begin{equation}
\label{eq:damped_poly_coord}
    z(r)=(r-r_{\rm e})\exp[-\beta_{2}(r-r_{\rm e})^2-\beta_{4}(r-r_{\rm e})^4],
\end{equation}
where $\beta_2$ and $\beta_4$ are damping constants.

\subsubsection{Empirical Rotational Born-Oppenheimer Breakdown, Spin-Spin, Spin-Rotational Curves}
\label{subsubsec:BobRot_SR_SS}

We fitted the phenomenological spin-spin (SS) couplings and the empirical spin-rotation (SR) couplings of the triplet \Xstate, \Astate, and \Bstate\ states to account for additional $\Omega$-splitting and to allow for additional variation of $J$, respectively, not described by the \ai\ model \citep{16YuLoTe.methods,93Kato.methods}. We also fited rotational Born-Oppenheimer breakdown (BOB) curves for all but the \Bstate\ state to correct for the electron un-coupling to the nuclear motion producing an additional $J^2$ dependence in the residuals of the rovibronic energies and which can be thought of as a correction to the position-dependent rotational mass. Some SS, SR, and BOB couplings are modelled using Eq.~(\ref{eq:damped_poly_coord}) and some using a \v{S}urkus polynomial expansion given by
\begin{equation}
    F(r)=(1-\xi_p)\sum_{i=0}^{N}a_{i}\xi_p^i+\xi_{p}a_{\infty}.
\end{equation}
This greatly enhanced the accuracy of the finalised spectroscopic model.

\subsection{Accuracy of the Refined Model}
\label{subsec:accuracy}

\begin{figure}
    \centering
    \includegraphics[width=0.99\linewidth]{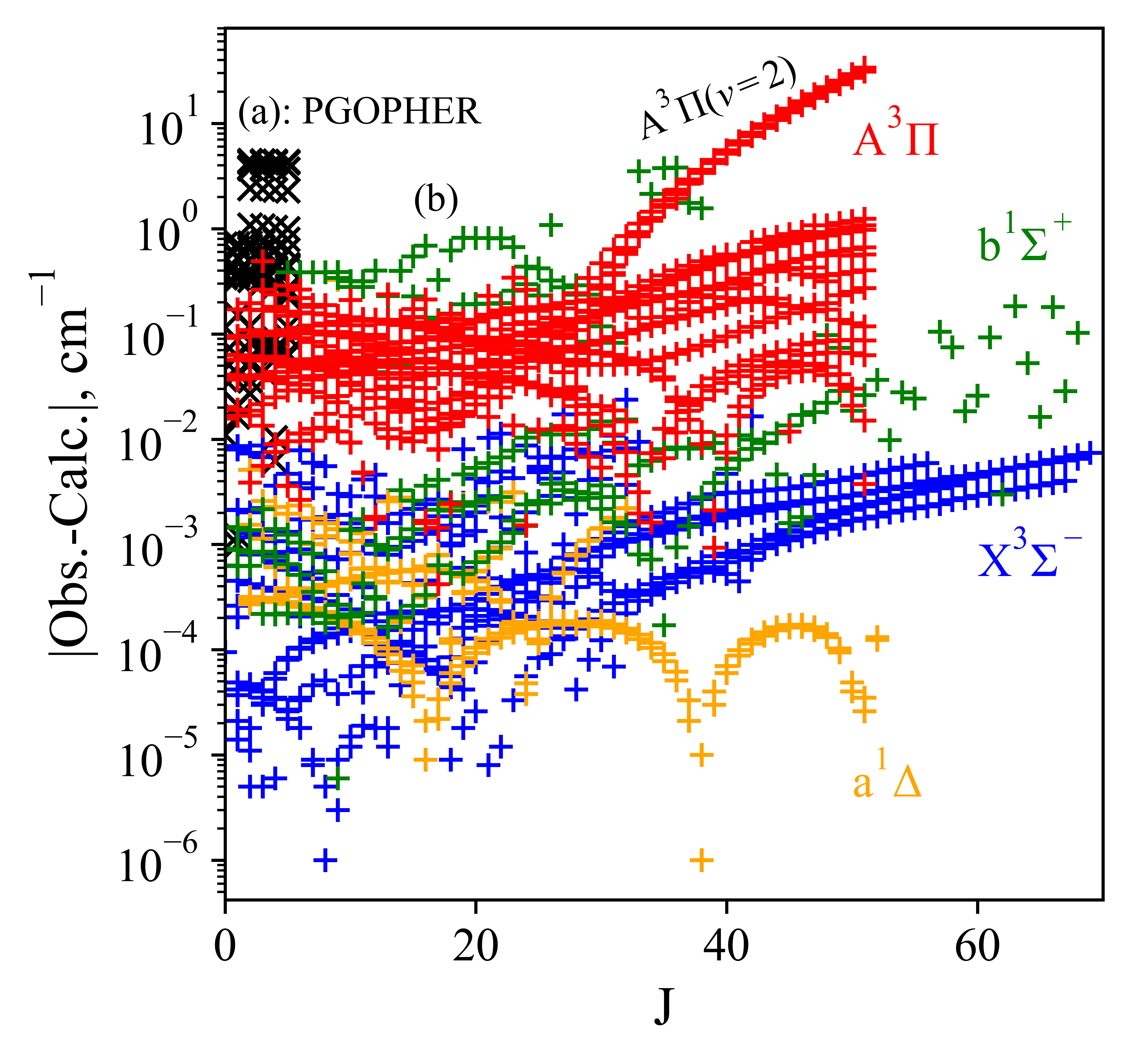}
    \caption{Visual representation of the difference between  our \marvel\ (Obs.) and the calculated (Calc.) energy levels  as a function of $J$ for the \Xstate, \astate, \bstate, and \Astate\ states. (a) The Black crosses compare computed PGOPHER energies to our calculated \duo\ ones; (b) high scatter present in the MARVELised \bstate($v=3,4$) levels; (\Astate($v=2$)) the Obs.-Calc. structure in the red points increases rapidly towards higher J's (see text).}
    \label{fig:O-C}
\end{figure}

Figure \ref{fig:O-C} illustrates the Observed minus Calculated (Obs.-Calc.) energy residuals as a function of the rotational quantum number $J$, and provides a metric on the accuracy of our model to reproduce our {\marvel}ised energies.  Most of the highly scattered energy levels have no combination differences with other sources, and so are effectively removed from the fit by setting their weight to $10^{-6}$. We fit 100$\%$ of 512 \Xstate\ $(J\leq69)$ energy levels with a total root-mean-square (rms) error of \wav{$3.13\times10^{-3}$}, 99$\%$ of 244 \astate\ $(J\leq 52)$ energy levels with a rms error of \wav{$1.08\times10^{-3}$}, 95$\%$ of 206 \bstate\ $(J \leq 64)$ energy levels with a total rms error of \wav{0.27}, and 78$\%$ of 1262 \Astate\ $(J \leq 34)$ energy levels with a  rms error of \wav{0.24}. The above rms errors were calculated after filtering outliers from our dataset which heavily influenced the rms, such as the scattered data of \bstate$(v=3,4)$, the \Astate$(v=2,J\geq35)$ states, and a single data point of \astate. \citet{99ElWexx.SO} provide rotational constants for the \Astate\ ($v=4-13$) states fitted to their 1+1 resonance enhanced multiphoton ionisation spectra which we used to compute low $J$ ($J\leq 5$) energies via \pgopher\ \cite{PGOPHER}. These energies to our knowledge are the only ones covering the highly excited vibronic states of \Astate\ and so we use them to constrain the \Astate\ potential up to its dissociation. The black points in figure \ref{fig:O-C} for $J\leq 5$ (see  label (a)) indicates  the vibrational dependence in the residuals of these \pgopher\ levels, which we manage to fit all within $\sim$\wav{4.5}. Doing this allowed for a more physical description of the effective position dependent correction to the rotational mass for the \Astate\ by constraining the potential gradient, and ultimately led to higher accuracy in the associated computed lifetimes (see Section \ref{subsec:int_scale}).

Generally, the \bstate\ data are of high quality and are reproduced by our model to within $\sim10^{-4}-10^{-3}$\wav{} with the exception of high scatter within the {\marvel}ised \bstate($v=3,4$) energies. This scatter can be seen in Fig.~\ref{fig:O-C} around the label (b).
One striking feature in Fig.~\ref{fig:O-C} is of the \Astate\ ($v=2$) Obs.-Calc. residuals, which, despite many attempts to model correctly, are poorly recovered in our model. We see a smooth, but rapid increase in the \marvel\ energies with $J$ of this band to $\sim$\wav{35} at $J=51$. We postulate this is because a dark state pushes these energies upwards, some candidates being the \cstate, \Aprimeprimestate, and \Aprimestate\ which cross through the \Astate\ potential, however all attempts to correctly position their PECs relative to the \Astate\ failed to reproduce this behaviour. No published data on the crossing states exists, but it is entirely possible that a correct description of these electronic potentials could resolve this issue, and some empirical data could be used to constrain their curves better than blindly varying their positions.

Some residual $J$-dependence can be seen for the \Xstate\ and \bstate\ states, where the former is due to $J$-dependent $\Omega$-splitting and parity splitting within the $|\Omega|=1$ levels, and the latter is due to vibrational dependence in the effective rotational centre not being fully accounted for. However, the residuals to the MARVEL energies are all  $\leq$\wav{$10^{-2}$} which means the model should extrapolate well to higher $J$.

One major problem faced during the refinement was with the spin-orbit splitting of the \Astate\ energies, where experiment \citep{69Colin.SO, 82Colin.SO} predicts regular $\Omega$ energy ordering, whereas multiple \ai\ calculations reveal the \brkt{\Astate}{{\rm SO}_z}{\Astate} SOC to have a negative phase, suggesting irregular $\Omega$ energy ordering. Analysis by \citet{69Colin.SO}  shows the $\Lambda$-doubling (LD)  to be $\sim$\wav{1.2} in the lowest energy $\Omega$ state with slight dependence on $J$, small doubling in the middle coupling which varies with $J(J+1)$ and zero splitting for the highest component state. To choose whether we adopt the irregular $\Omega$ energy ordering suggested by \ai\ calculations or the experimental assignment with a change in phase of the \brkt{\Astate}{{\rm SO}_z}{\Astate} SOC we studied the $e/f$ parity splitting of the \Astate\ energies since this would confirm what $\Lambda$-doubling matrix elements to adopt. We saw that the slightly $J$-dependant $\sim$\wav{1.2} parity splitting in the lowest energy state could only be resolved via the following $\Lambda$-doubling element
\begin{equation}
\label{eq:LD}
    \hat{\mathcal{H}}_{\rm LD}=\frac{1}{2}\alpha_{opq}^{\rm LD}(r)(\hat{S}^2_{+}+\hat{S}^2_{-})
\end{equation}
with the $\Lambda$-doubling constant being consistent with the \citet{79BrMexx.methods} convention $\alpha^{\rm LD}_{opq}=o^{\rm LD}+p^{\rm LD}+q^{\rm LD}$ for a regular $\Omega$ energy ordering, i.e. the splitting is between states of $\Delta\Sigma=\pm2$ which is only possible for a triplet state if $\Omega=0$. If one adopted an irregular assignment scheme with $\Omega=2$ for the lowest energy state, then one could not correctly model the $J$-dependence of the splitting since the element in Eq.~ \eqref{eq:LD} would be zero. We thus adopted the experimental assignment and changed the sign of our \brkt{\Astate}{{\rm SO}_z}{\Astate} SOC which should not break the phase consistency of the model since it is a diagonal coupling.

\subsection{Dipole Moment Curves}
\label{subsec:DMCs}
\begin{figure}
    \centering
    \includegraphics[width=\linewidth]{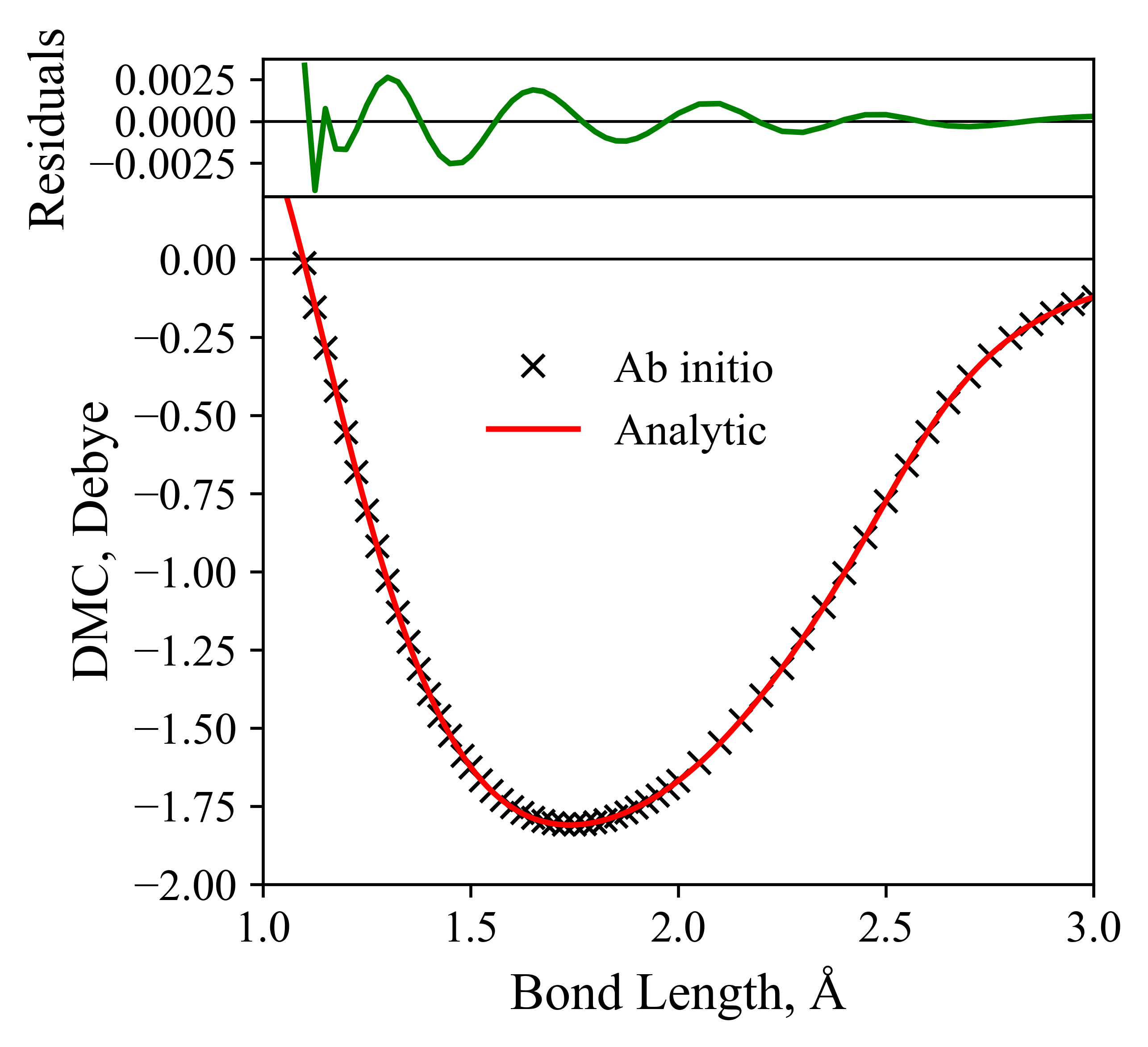}
    \caption{The \ai\ \Xstate\ DMC provided by \citet{22BeJoLi.SO}, computed at an ic-MRCI+Q level of theory with full relativistic corrections using aug-cc-pCV6Z-X2C basis sets, is shown (black crosses) superimposed with our fitted analytical form using Eq. \ref{eq:irreg_cheby} (red line). The residuals to the \ai\ DMC of our fit are shown in the top panel (green line).}
    \label{fig:irreg_cheby_fit}
\end{figure}

We use the accurate \ai\ ground state dipole moment function from \citet{22BeJoLi.SO} who calculate with an ic-MRCI+Q level of theory including the Davidson corrections, scalar relativistic contributions using the exact 2-component (X2C) relativistic Hamiltonian, and aug-cc-pCV6Z-X2C basis sets. All other DMCs are computed at a level of theory described in Section \ref{subsec:ai_model}. Within nuclear motion and intensity calculations, these dipoles are originally represented as a grid of \ai\ points on the \duo\ defined grid, however one sees a flattening of both the IR {\Xstate}--{\Xstate} band spectrum and its variation of TDMC with vibrational excitation. The source of this nonphysical flattening has been discussed by \citet{15MeMeSt.CO} and \citet{16MeMeSt,22MeUs.CO} who identify numerical noise as the culprit. This noise comes from \duo\ interpolating the given \molpro\ dipole grid points onto the \duo\ defined grid. One can try to increase the precision of their transition moments from double to quadruple precision, but this seldom fixes the problem with any appreciable magnitude. The most effective method found is to represent the input dipole moments analytically. We chose to represent our \Xstate\ DMC using the 'irregular DMC' proposed in \citet{22MeUs.CO} which takes the form
\begin{equation}
\label{eq:irreg_cheby}
    \mathcal{D}_{\rm irreg}(r)= \chi(r;c_2,...,c_6)\sum_{i=0}^6 b_iT_i(z(r))  
\end{equation}
Where $T_i$ are Chebyshev polynomials of the first kind, $b_i$ are summation coefficients to be fit, $z(r)$ is a reduced variable in bond length similar to the damped polynomial coordinate in Eq.~\eqref{eq:damped_poly_coord} and is given by,
\begin{equation}
    z(r)=1-2\, e^{-c_1r},
\end{equation}
which maps the $r\in[0,\infty]$ interval to the $z\in[-1,1]$ reduced interval, and finally $\chi(r;c_2,...,c_6)$ is an $r$-dependent term parametrically dependent on 5 $c_k$ parameters to be fitted and is given by 
\begin{equation*}
\chi(r;c_2,...,c_6)\frac{(1-e^{-c_2r})^3}{\sqrt{(r^2-c_3^2)^2+c_4^2}\sqrt{(r^2-c_5^2)^2+c_6^2}}.
\end{equation*}
Our fitted \Xstate\ DMC is illustrated with its residual to the \ai\ DMC in Fig.~\ref{fig:irreg_cheby_fit}. 

The irregular DMC has the desirable properties of quickly converging to the correct long-range limit, having enough parameters (13) to ensure an accurate description of the full range in bond length with minimal local oscillations, and provide a straight Normal Intensity Distribution Law (NIDL) \citep{22MeUs.CO,12Medvedev,15MeMeSt.CO}. 

\begin{figure*}
    \centering
    \includegraphics[width=0.98\linewidth]{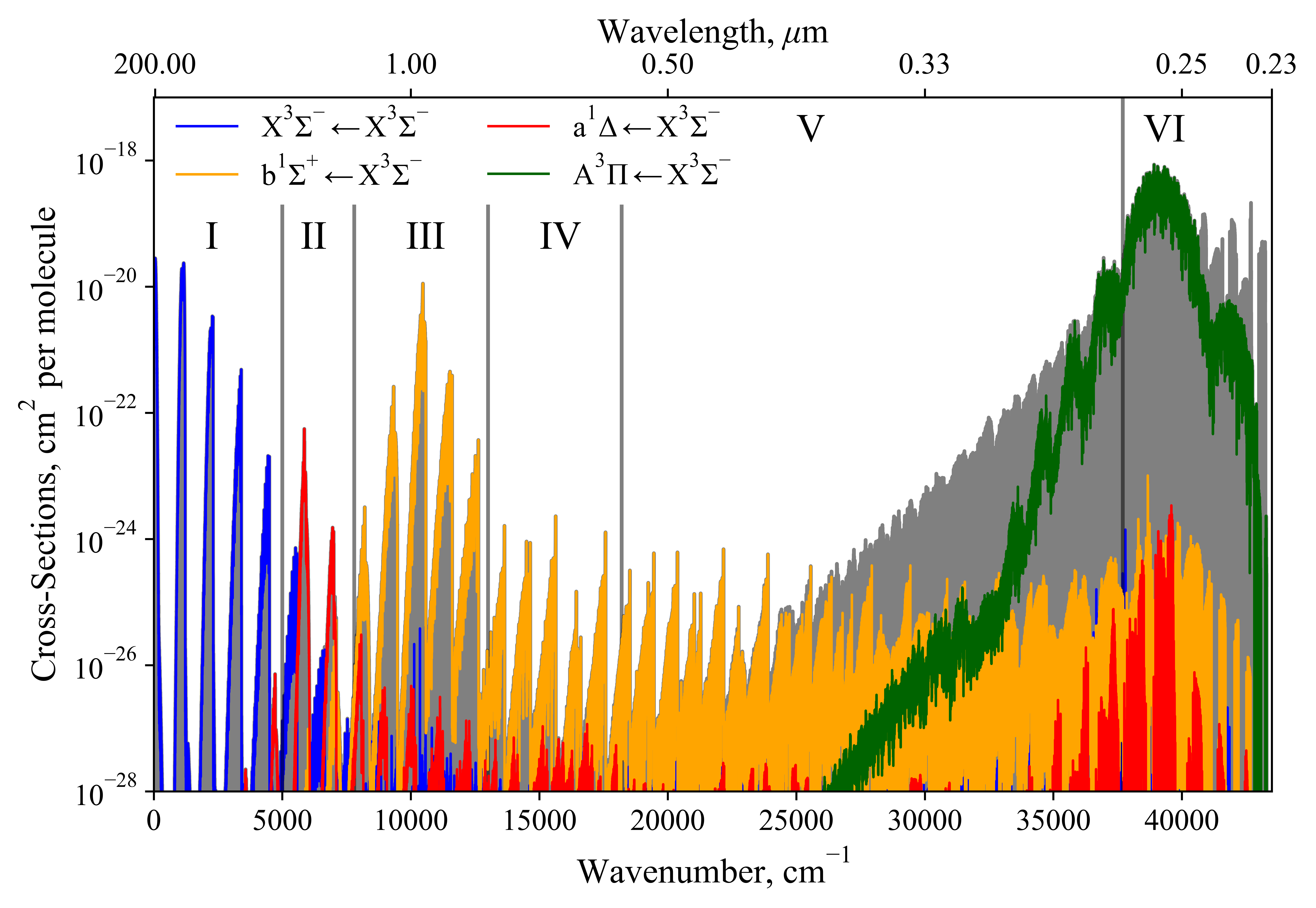}
    \caption{Dipole allowed and forbidden components of the absorption spectrum simulated with our semi-empirical model at 1000~K connecting \Xstate\ with \Xstate, \astate, \bstate, and \Astate. Regions of spectral importance are marked with roman numerals and are detailed in the text. The grey shaded region marks the total SO opacity computed with our model at 1000 K.}
    \label{fig:total spec}
\end{figure*}
\begin{figure}
    \centering
    \includegraphics[width=0.98\linewidth]{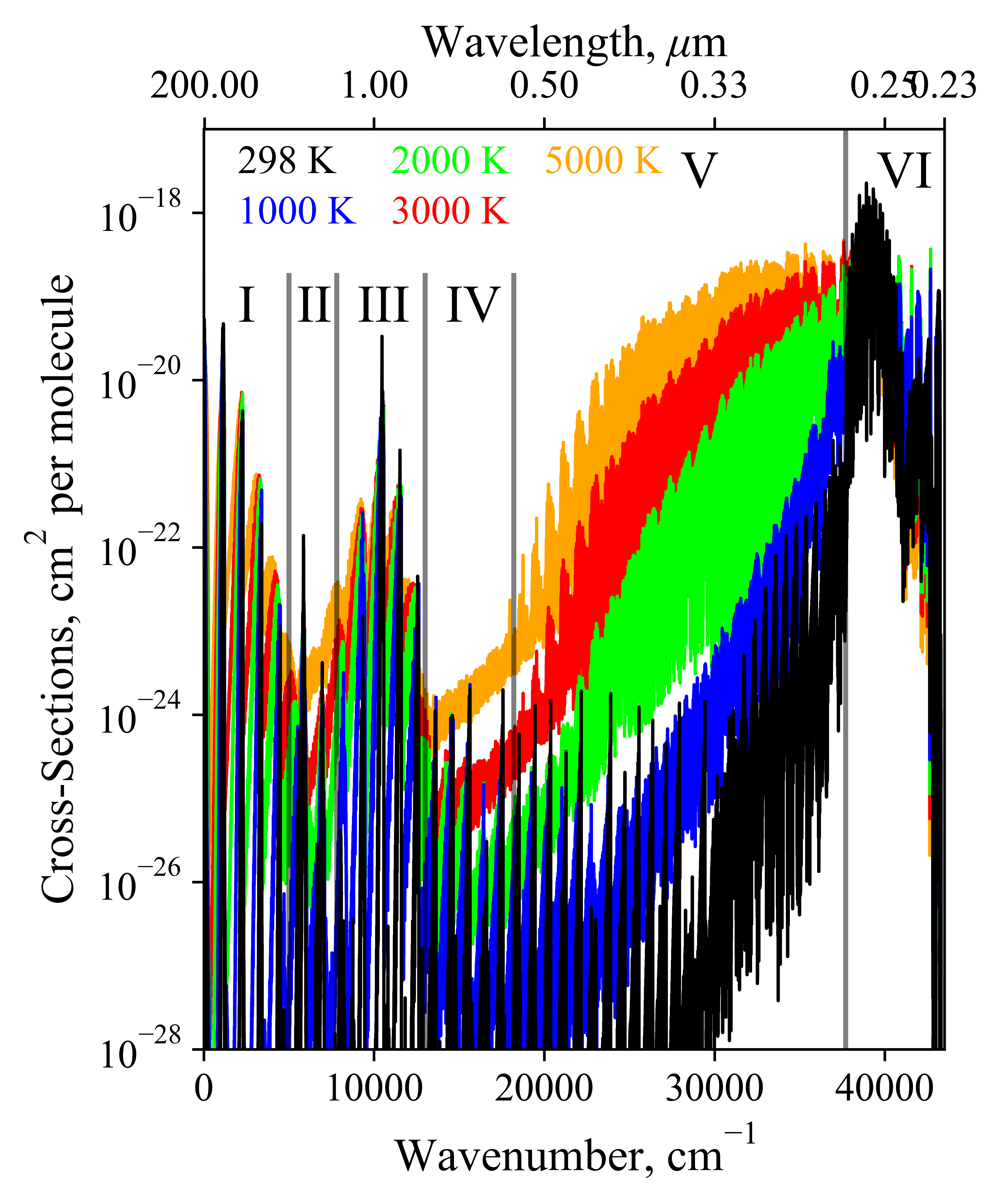}
    \caption{The total absorption spectrum of SO simulated with our semi-empirical model for different temperatures ranging from 298 K to 5000 K. We see the intensity deviation is greatest in region $V$ around \wav{18~000-35~000} where the \Bstate$\leftarrow$\Xstate band begins to dominate opacity.}
    \label{fig:temp spec}
\end{figure}

\section{Line List}
\label{sec:Line List}

We produce a semi-empirical rovibronic line list \LineList\ for $^{32}$S$^{16}$O  covering the \Xstate, \astate, \bstate\ and \Astate\ electronic states, where a system involving couplings between \Xstate, \astate, \bstate, \Astate, \Bstate, \Aprimeprimestate, \Aprimestate, and \estate\ defines our spectroscopic model. The \textsc{SOLIS} line list covers  wavelengths  down to to 222.22~nm. Illustrations of the spectra simulated with the new line list are presented in Figs.~(\ref{fig:total spec},\ref{fig:temp spec}).

For nuclear motion calculations a vibrational sinc-DVR basis set was defined for a grid of 301 internuclear geometries in the range {0.6}--{6.0}~{\AA}. We select  58 vibrational wavefunctions for the \Xstate, \astate, \bstate, \Astate, \Bstate, \Aprimeprimestate, \Aprimestate, and \estate\ states to form the contracted vibronic basis. In total 7~008~190 Einstein A coefficients between 84~114 bound rovibronic states were computed with a maximum total rotational quantum number $J_\text{max}$ = 250.

\begin{table*}
\centering
\caption{ Extract from the states file of the line list for SO.}
%\tt
\label{tab:states}
{\tt  \begin{tabular}{rrrrrrcclrrrrcrr} \hline \hline
%  &  $\tau$ & $g$&  & Label & Calc. 
$i$ & Energy (\cm) & $g_i$ & $J$ & unc &  $\tau$ &  \multicolumn{2}{c}{Parity} 	& State	& $v$	&${\Lambda}$ &	${\Sigma}$ & $\Omega$ & Ma/Ca & Energy (\cm) \\ 
\hline
733 & 12277.658473 & 5 & 2 & 0.302298 & 0.068323 & +  & e & a1Delta & 6 & 2 & 0 & 2 & Ca & 12277.658473 \\ 
734 & 12576.55717 & 5 & 2 & 0.01039 & 0.0067518 & +   & e & b1Sig+ & 2 & 0 & 0 & 0 & Ma & 12576.558535 \\ 
735 & 12824.746272 & 5 & 2 & 0.603962 & 0.025301 & +  & e & X3Sig- & 12 & 0 & 0 & 0 & Ca & 12824.746272 \\ 
736 & 12836.684546 & 5 & 2 & 0.603962 & 0.025272 & +  & e & X3Sig- & 12 & 0 & 1 & 1 & Ca & 12836.684546 \\ 
737 & 13297.933546 & 5 & 2 & 0.352298 & 0.059146 & +  & e & a1Delta & 7 & 2 & 0 & 2 & Ca & 13297.933546 \\ 
738 & 13602.425655 & 5 & 2 & 0.165193 & 0.0069414 & + & e & b1Sig+ & 3 & 0 & 0 & 0 & PS & 13601.834707 \\ 
739 & 13810.582705 & 5 & 2 & 0.653962 & 0.023019 & +  & e & X3Sig- & 13 & 0 & 0 & 0 & Ca & 13810.582705 \\ 
740 & 13822.57783 & 5 & 2 & 0.653962 & 0.022993 & +   & e & X3Sig- & 13 & 0 & 1 & 1 & Ca & 13822.57783 \\ 
\hline
\hline
%%%%
%%%%
\end{tabular}}
\mbox{}\\
{\flushleft
$i$:   State counting number.     \\
$\tilde{E}$: State energy term values in \cm, MARVEL or Calculated (\textsc{Duo}). \\
$g_i$:  Total statistical weight, equal to ${g_{\rm ns}(2J + 1)}$.     \\
$J$: Total angular momentum.\\
unc: Uncertainty, \cm.\\
$\tau$: Lifetime (s$^{-1}$).\\
$+/-$:   Total parity. \\
$e/f$:   Rotationless parity. \\
State: Electronic state.\\
$v$:   State vibrational quantum number. \\
$\Lambda$:  Projection of the electronic angular momentum. \\
$\Sigma$:   Projection of the electronic spin. \\
$\Omega$:   Projection of the total angular momentum, $\Omega=\Lambda+\Sigma$. \\
Label: `Ma' is for MARVEL, `Ca' is for Calculated, and 'PS' is for predicted shift. \\
Energy: State energy term values in \cm, Calculated (\textsc{Duo}). \\
}
\end{table*}

\begin{table}
\centering
\caption{Extract from the transitions file of the line list for SO. }
\tt
\label{tab:trans}
\centering
\begin{tabular}{rrrr} \hline\hline
\multicolumn{1}{c}{$f$}	&	\multicolumn{1}{c}{$i$}	& \multicolumn{1}{c}{$A_{fi}$ (s$^{-1}$)}	&\multicolumn{1}{c}{$\tilde{\nu}_{fi}$} \\ \hline
       37557    &    36527  & 2.7817E-01    &     5199.704942 \\
       37204    &    36852  & 2.7817E-01    &     5199.704945 \\
       32098    &    32422  & 1.2080E+00    &     5199.713048 \\
       21055    &    22048  & 3.3851E-06    &     5199.718029 \\
       60350    &    61047  & 2.8777E-04    &     5199.728151 \\
       45755    &    46561  & 4.3835E-01    &     5199.728902 \\
    \hline\hline
\end{tabular} \\ \vspace{2mm}
\rm
\noindent
$f$: Upper  state counting number;\\
$i$:  Lower  state counting number; \\
$A_{fi}$:  Einstein-$A$ coefficient in s$^{-1}$; \\
$\tilde{\nu}_{fi}$: transition wavenumber in \cm.\\
\end{table}

The PEC of the \Astate\ state implies that predissociative and continuum states  should exist for the region above dissociation. To this end, these states have been removed from the line list through checking the character of the wavefunctions at the `right' simulation border $r_{\rm max}$ in our \duo\ model where unbound states tend to oscillate at $r\to \infty$  with a non-zero density around  $r_{\rm max}$ \citep{22YuNoAz}. The line list we present therefore only contains bound to bound transitions only.

The calculated energies in the \texttt{.states} file are `MARVELised' which involves replacing them with the MARVEL ones. For levels that are not covered by the MARVEL SN, the predicted shift method of  \citet{jt835} was used to MARVELise them. Predicted shifts work by fitting the Obs.-Calc. trends as functions of $J$ for each `state', $v$ and $\Omega$ energy band to then interpolate gaps within the MARVEL network or extrapolating to higher $J$. 

The \LineList\ line list is available in the ExoMol database (\href{https://www.exomol.com/}{https://www.exomol.com/}) in the form of a  States (\texttt{.states}) and Transition (\texttt{.trans}) files, with extracts shown in Tables \ref{tab:states} and \ref{tab:trans} respectively. Uncertainties for the energy levels where either taken directly as the MARVEL ones where available, or otherwise computed using the following empirical formulae
\begin{align}
\label{e:unc:form}
    \sigma({\rm state},J,v) = \Delta T + \Delta\omega\, v + \Delta B\, J(J+1),
\end{align}
where $\sigma$ is the energy uncertainty for a given state and $\Delta T$, $\Delta \omega$, $\Delta B$ are state dependent parameters given in Table \ref{tab:unc_const}. $\Delta T$ were found by taking twice the standard deviation of the total Obs.-Calc. of each electronic state (see Fig. \ref{fig:O-C}) after outliers where removed by selecting states outside of this two standard-deviation threshold, where the standard deviation was computed again.
\begin{table}
\centering
\caption{ State dependent parameters (in cm$^{-1}$) of Eq.~\eqref{e:unc:form} used to estimated uncertainties for the calculated states of \SO\ where MARVEL uncertainties were not available.}
\label{tab:unc_const}
\begin{tabular}{lrrrr} 
\hline \hline
State & $\Delta T$ & $\Delta \omega$ & $\Delta B$  \\ 
\hline
\Xstate\ & 0.003363 & 0.05 & 0.0001 \\
\astate\ & 0.001698 	&0.05 & 0.0001 \\
\bstate\ & 0.368965 	&0.05 & 0.0001  \\
\Astate\ & 2.835039 	&0.05 &	0.0001 \\
    \hline\hline
\end{tabular} 
\end{table}

\subsection{Intensity Scaling: Dipoles and Lifetimes}
\label{subsec:int_scale}

There are only a few recorded experimental values for electric dipole moments, lifetimes, and no direct intensity measurements which can be used to constrain our \ai\ dipoles. Lifetimes are useful to constrain dipole moments via the relation
\begin{equation}
    \frac{1}{\tau_u} =\sum_l A_{ul}  \propto |\brkteq{u}{\mu_{\sigma}}{\tilde{l}}|^2,
\end{equation}
where $\ket{\tilde{l}}$ is the dominant ro-vibronic state contributing to the lifetime of the level $\ket{u}$, $\sigma=0,\pm1$ denotes a tensorial dipole component. 
So a scaling in lifetime $\tilde{\tau}=\xi\tau$ would correspond to an approximate scaling in dipole moment to the dominant lower state of $1/\sqrt{\xi}$.

Previous Stark measurements \citep{64PoLixx.SO, 92LoSuOg.SO} have determined the ground state dipole to be $\mu_X^{0}=$ 1.55(2)~D \citep{64PoLixx.SO} and 1.52(2)~D \citep{92LoSuOg.SO}, slightly smaller than our computed vibrational transition moment of $|\bra{\Xstate,v=0}\mu_{0}\ket{\Xstate,v=0}| =1.588$ D. We scale our \Xstate\ dipole to the value of 1.535~D averaged from the two Stark measurements, which we note is the dipole moment adopted by CDMS \citep{CDMS} where the same averaging was done. \citet{83WiFiWi.SO} measure the radiative lifetime of the \bstate$(v=0)$ state through time-resolved measurements of the \band{\bstate}{\Xstate} emission band and provide a lifetime of $\tau=6.8\pm 0.4$~ms. To achieve this lifetime, we scale our \brkt{\bstate}{\mu_{0}}{\bstate} dipole by a factor of $0.7401$.

\citet{70Saito.SO} determines the \astate\ dipole moment to be $1.336\pm0.045$ D through Stark measurements, larger than our computed transition moment $|\bra{\astate,v=0}\mu_{0}\ket{\astate,v=0}| =1.184$ D. We scale our \astate\ dipole by a factor of 1.1282 to reproduce the measured transition moment.

\begin{figure}
    \centering
    \includegraphics[width=\linewidth]{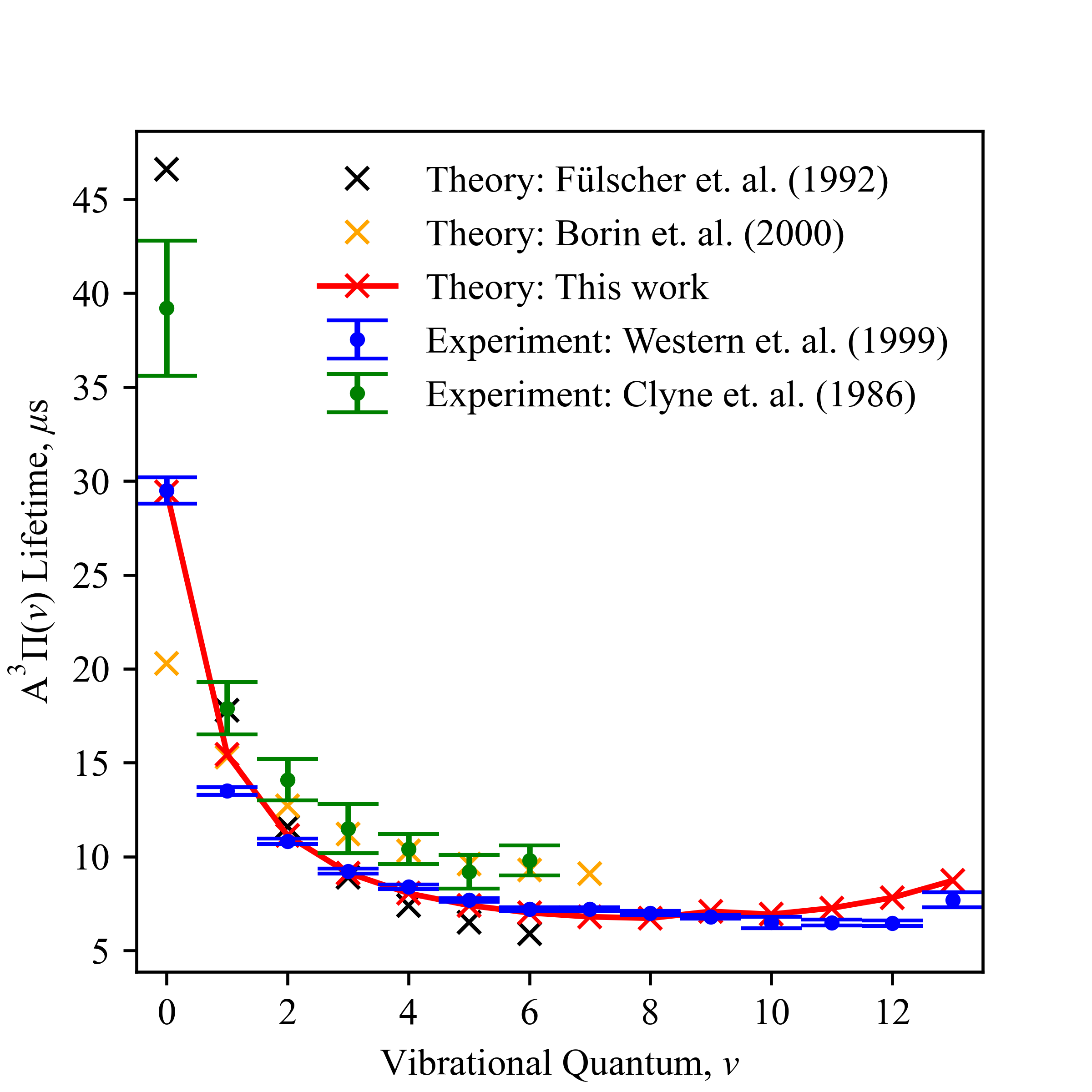}
    \caption{The\Astate\ state lifetimes as a function of $v^{\prime}$ are shown from experimental \citep{99ElWexx.SO,86ClTexx.SO} and theoretical \citep{00BoOrxx.SO, 92FuJaRo.SO} sources with our computed lifetimes overlaid in red.}
    \label{fig:A LT}
\end{figure}

Radiative lifetimes of the \Astate\ state for $v^{\prime}=0-13$ were measured by \citet{99ElWexx.SO} by laser induced fluorescence and \citet{82ClLixx.SO} for $v^{\prime}=0-6$. Figure \ref{fig:A LT} shows the experimentally determined lifetimes as a function of $v^\prime$ with the theoretically predicted values by \citet{00BoOrxx.SO} and \citet{92FuJaRo.SO} and those computed by our model superimposed in red. Since \citet{99ElWexx.SO} quote their lifetimes to a lower uncertainty and for a large vibrational coverage we chose to model their lifetimes. Modelling these lifetimes proved to be difficult, the characteristic sharp drop in lifetime from $v^{\prime}=0$ and $v^{\prime}=1$ was very sensitive to multiple factors: (1) the position of the \brkt{\Astate}{\mu_{\pm1}}{\Xstate} dipole relative to the respective PECs; (2) the crossing point of the dipole with zero; (3) the local gradient of the dipole around the zero crossing point. Initial attempts to reproduce the experimental lifetimes were made using our \ai\ dipole, various \ai\ dipoles from the literature \citep{92FuJaRo.SO,19SaNaxx.SO,19FeZhxx.SO}, and the empirical dipole from \citet{99ElWexx.SO} in our model all failed to produce lifetimes that agreed with experiment. 

Firstly, we were able to reproduce the lifetimes of \citet{99ElWexx.SO} using a linear dipole function. Albeit being nonphysical, it provided important constraints on the short range position of and its gradient around the equilibrium geometry.  
We then combined it with the MRCI-F12+Q/aug-cc-pV(5+d)Z dipole computed by \citet{19SaNaxx.SO} at larger values of $r$ into a single smooth curve. Despite a slight change in the shape of the \ai\ dipole,  Fig.~\ref{fig:A LT} shows that our semi-empirically fitted \brkt{\Astate}{\mu_{\pm1}}{\Xstate} dipole is much closer to the \ai\ curve than the empirically fitted dipole by \citet{99ElWexx.SO}.

\subsection{Partition Function}
\label{subsec:PF}

\begin{figure}
    \centering
    \includegraphics[width=\linewidth]{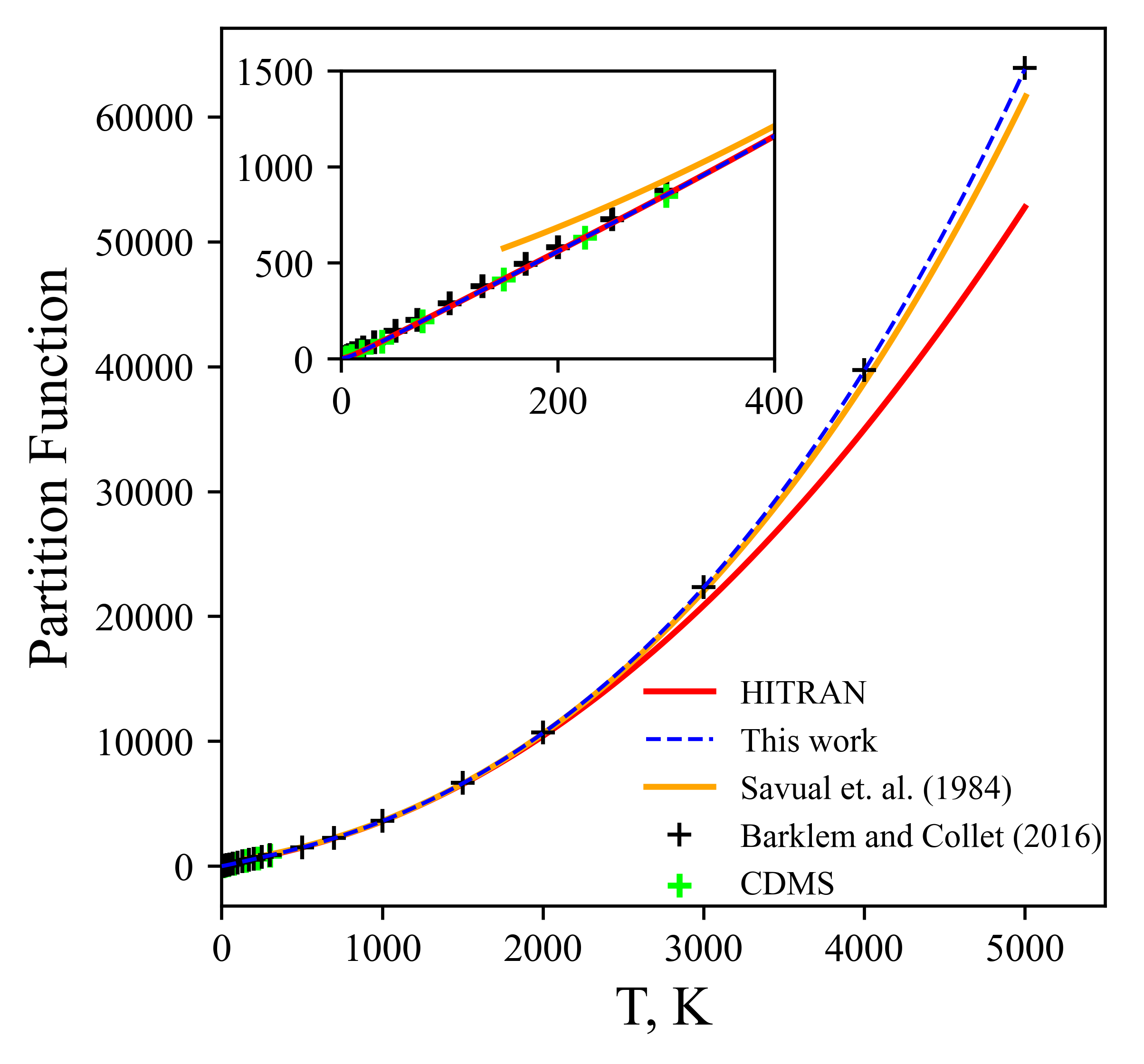}
    \caption{Comparison between our partition function and those produced by HITRAN \citep{jt692}, \citet{84SaTaxx.partfunc}, \citet{16BaCoxx.partfunc} and CDMS \citep{CDMS}.}
    \label{fig:PF}
\end{figure}

We compute the molecular partition function (PF) for SO from our semi-empirical line list using
\begin{equation}
    Q(T)=\sum_{i}g_i^{\rm tot}e^{-\frac{c_2 \tilde{E}_i}{T}}
\end{equation}
where $c_2$ is the second radiation constant, $\tilde{E}_i$ is the rovibronic energy term value in wavenumbers, $g_i^{\rm tot}=g_{\rm ns}(2J_i+1)$ is the total state degeneracy which includes the nuclear weight spin-statistic $g_{\rm ns}$ ($g_{\rm ns}$ = 1 for $^{32}$S$^{16}$O) where we use a 1 K temperature step. Figure \ref{fig:PF} compares our computed PF to the PFs of \citet{84SaTaxx.partfunc}, \citet{16BaCoxx.partfunc}, CDMS \citep{CDMS}, and \hitran\ \citep{jt692} who compute their PF from the line lists produced by \citet{21BeJoLi.SO, 22BeJoLi.SO}. As the nuclear spin degeneracy is one, no PFs  need to be scaled  to the physics convention of nuclear statistical weights, which ExoMol uses. Figure \ref{fig:PF} shows that all PFs agree for $500\lesssim T\lesssim 2000$ K; for all temperatures our computed PF continues to agree with that of \citet{16BaCoxx.partfunc}, where our computed PF is generally lower than theirs up to $0.1\%$ at 5000 K; the CDMS PF agrees to within $1\%$ of all PFs up to it cutoff at 300 K; \citet{84SaTaxx.partfunc} is larger than the rest of the PFs at low temperatures up to 500 K and at 5000 K their PF is $3.4$\% lower than our computed PF; the \hitran\ PF begins to deviate from the other PFs from about 2000 K, where at 5000 K it is lower than ours by $17\%$; this behaviour of \hitran\ PFs has been noted previously \citep{jt899}.

\subsection{SO opacities}

We follow the ExoMolOP procedure of \citet{20ChRoYu} and generate molecular opacitites for SO using the SOLIS line list for four exoplanetary atmosphere retrieval codes ARCiS \citep{ARCiS}, TauREx \citep{TauRex3}, NEMESIS \citep{NEMESIS} and petitRADTRANS \citep{19MoWaBo.petitRADTRANS} on an extensive  grid of temperatures and pressures. The opacities are provided as part of the SO ExoMol data set  at www.exomol.com.

\subsection{Simulated spectra}
\label{subsec:simspec}
Program \exocross\ \citep{ExoCross} was used to simulate rovibronic absorption  spectra as a function of temperature using \LineList.
Figure \ref{fig:total spec} illustrates the dipole allowed and forbidden electronic bands connecting the \Xstate\ state to \Xstate, \astate, \bstate, and \Astate\ which are shown as different colours and the total computed SO opacity is shown in grey. Here we simulate lines with a Gaussian line profile of HWHM \wav{0.6}. The forbidden band intensities are stolen through mixing of the electronic wavefunctions through couplings such as SOCs, DMCs, and EAMCs resulting in non-zero  dipole matrix elements, which we note provides a stronger mechanism here than their corresponding magnetic dipole or electric quadrupole couplings. Figure \ref{fig:temp spec} shows the temperature variation of the simulated total SO opacity which has a strong effect on the UV/Vis cross-sections. The greatest temperature variation can be seen in the \wav{18~000-35~000} region (V) where the \band{\Xstate}{\Bstate} band begins to dominate opacity. Here we simulate lines with a Gaussian line profile of HWHM \wav{0.6}. It is clear the IR/NIR spectrum is largely unaffected by the increase of temperature except from the expected rotational broadening. Below we comment on the spectral regions marked by I-VI illustrated in Fig.~\ref{fig:total spec}.

(I) The IR $\sim$\wav{0--5000} region is dominated by the \Xstate$\leftarrow{}$\Xstate\ electronic band peaking at $\sim3\times10^{-20}$ cm$^2$ per molecule.

(II) The $\sim$\wav{5000--7800} NIR region shows strong  \astate$\leftarrow{}$\Xstate\ band features, even for room temperature spectra, but \Xstate$\leftarrow{}$\Xstate\ lines are expected to be still observable here.

(III) The $\sim$\wav{7800--13000} NIR region is dominated by strong \bstate$\leftarrow{}$\Xstate\ band absorption for all temperatures, and is almost as strong as the dipole allowed \Xstate$\leftarrow{}$\Xstate\ band spectrum because of large intensity stealing mechanism facilitated through the strong \brkt{\bstate}{\mu_z}{\Xstate} SOC, see the Appendix.

(IV) The Vis $\sim$\wav{13000--18200} region shows a flat feature due to \Astate$\leftarrow{}$\Xstate\ band absorption which becomes prominent for temperatures above 3000~K. However, since we omit the \Cstate\ state in our spectroscopic model we do not compute \Cstate$\leftarrow{}$\Aprimestate\ and \Cstate$\leftarrow{}$\Aprimeprimestate\ band intensities which we previously predicted  to be strong in this region \citep{22BrYuKi.SO}. 

(V) The Vis/UV $\sim$\wav{18200--37700} region is largely uncovered at high accuracy by our spectroscopic model since it is dominated by the \Bstate$\leftarrow{}$\Xstate\ and lesser \Cstate$\leftarrow{}$\Xstate\ electronic bands which become major sources of SO opacity for temperatures above 1000~K. We are currently working on the UV SO line list for a future study which will accurately cover this region. However, for lower temperatures the \Astate$\leftarrow{}$\Xstate\ and \bstate$\leftarrow{}$\Xstate\ bands become more important which we recover accurately in this study.

(VI) The UV $\sim$\wav{37700--43500} region has a strong \Astate$\leftarrow{}$\Xstate\ band feature which should be observable at all temperatures.

\subsection{Comparisons to experimental spectra}
\label{subsec:comp_to_exp_spec}

There are few recorded experimental spectra of \SO\ with large coverage and almost none with absolute intensity measurements, relative intensities are usually provided \citep{02ChWaLi.SO, 87BuLoHa.SO, 99SeFiRa.SO, 05WaTaHa.SO}. The only study to our knowledge that provides measured absolute intensities is the recent study by \citet{22HeStLy.SO} on the \Astate$-$\Xstate\ band, which we compare to (also \Bstate$-$\Xstate\ and \Cstate$-$\Xstate bands which we do not compare to). 

The forbidden band intensities we compute here are through the intensity stealing mechanism which works through mixing of electronic state wavefunctions through couplings such as SOCs. We do not compute magnetic dipole intensities, which are much weaker for the bands of interest than the redistributed intensities which we compute. For example, the diagonal \brkt{\bstate}{\mu_{z}}{\bstate} and \brkt{\Xstate}{\mu_{z}}{\Xstate} dipoles produce \bstate$\leftarrow$\Xstate\ band intensities a factor of $\sim 10-1~000$ times stronger at the band peak than the corresponding magnetic dipole intensities. Therefore, we omit magnetic dipole transitions from our line list. An example of the intensity stealing mechanism is given in Appendix.

\begin{figure}
    \centering
     \includegraphics[width=\linewidth]{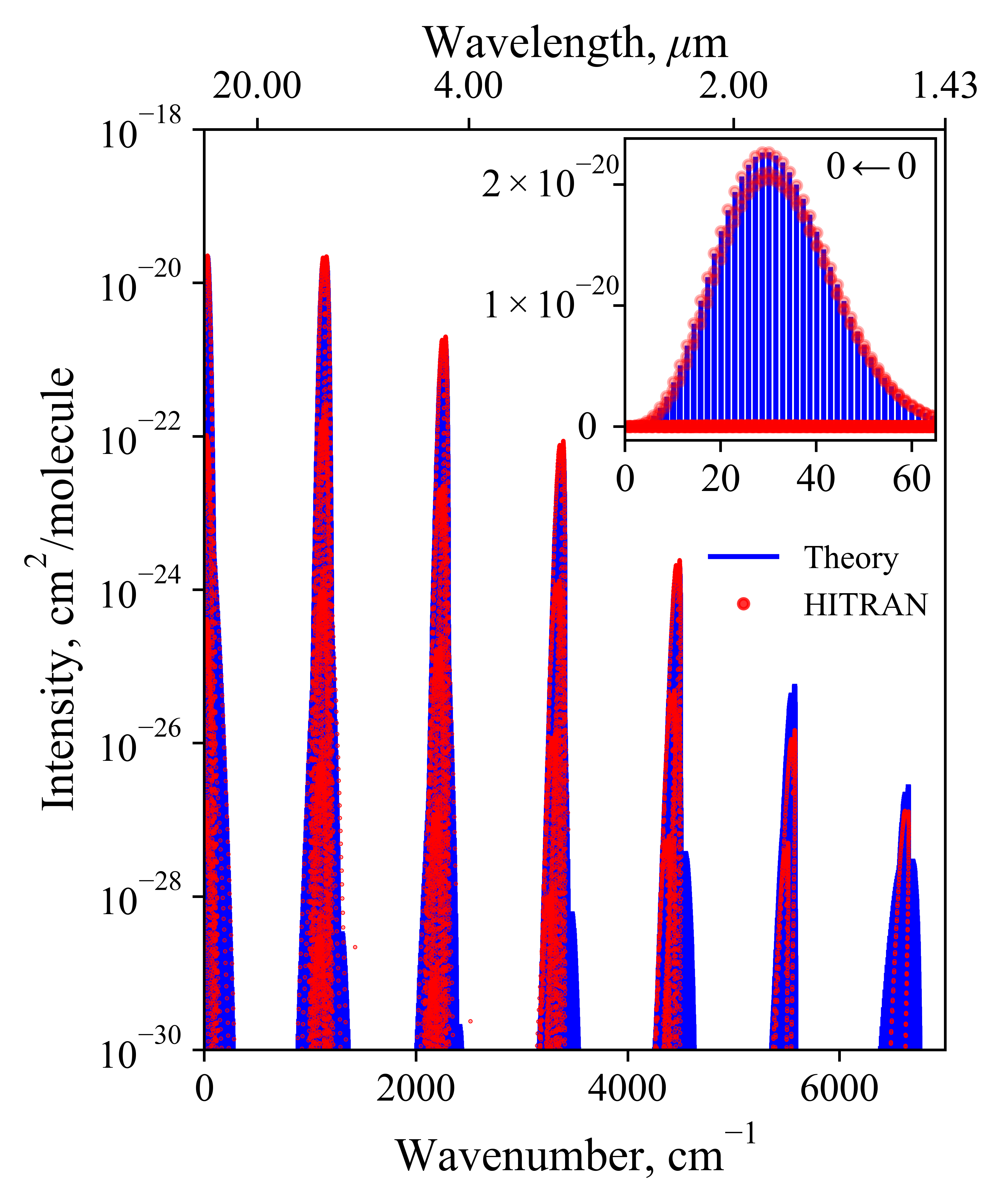}
    \caption{Comparison between the theoretical and HITRAN \band{\Xstate}{\Xstate} rovibrational band for  \wav{0-7000}. We simulate the spectra using a temperature of 296 K and scale the intensities by the fractional isotopologue abundance of 0.9479 \citep{jt857}.}
    \label{fig:X_HITRAN}
\end{figure}
\begin{figure}
    \centering
    \includegraphics[width=\linewidth]{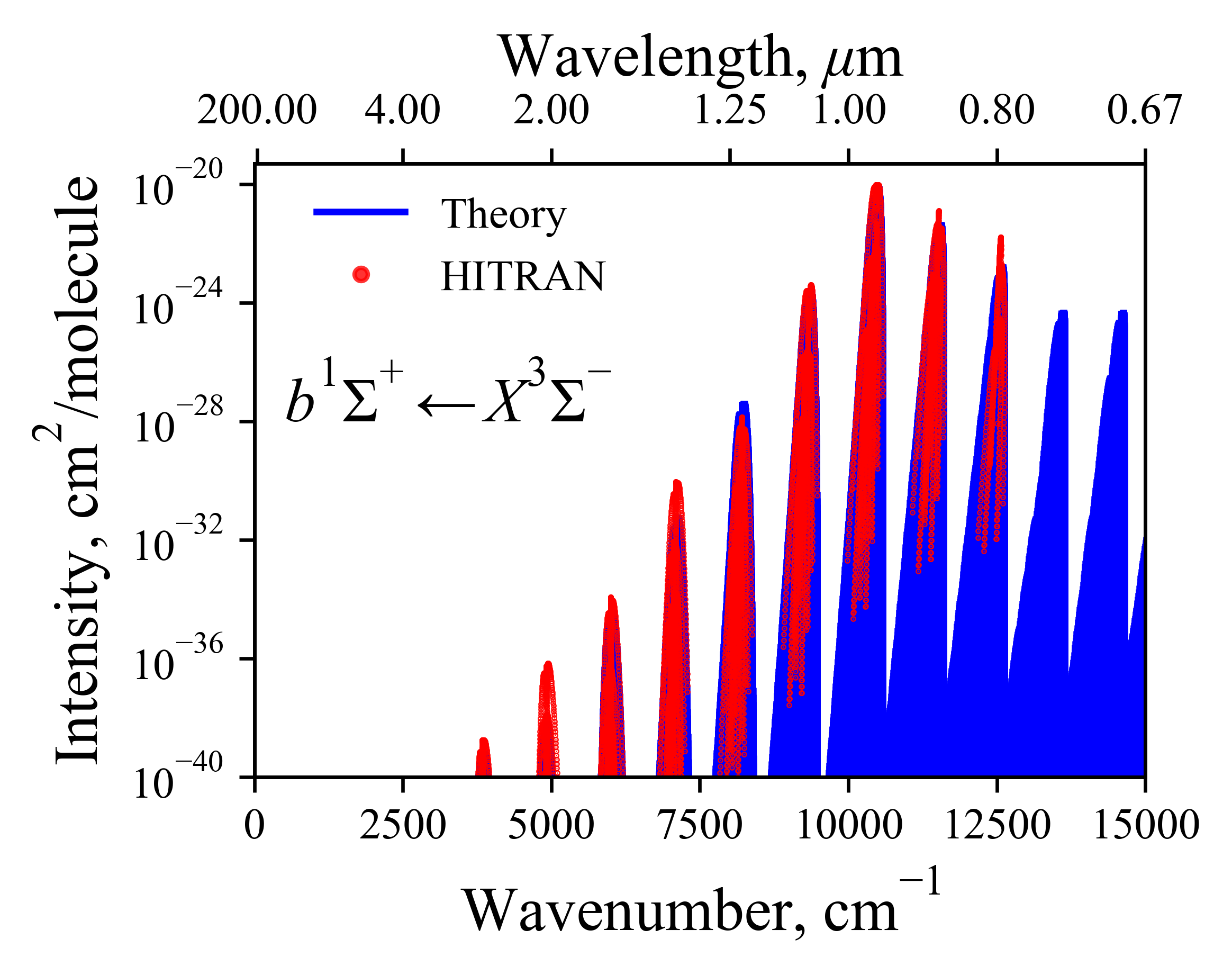}
     \includegraphics[width=\linewidth]{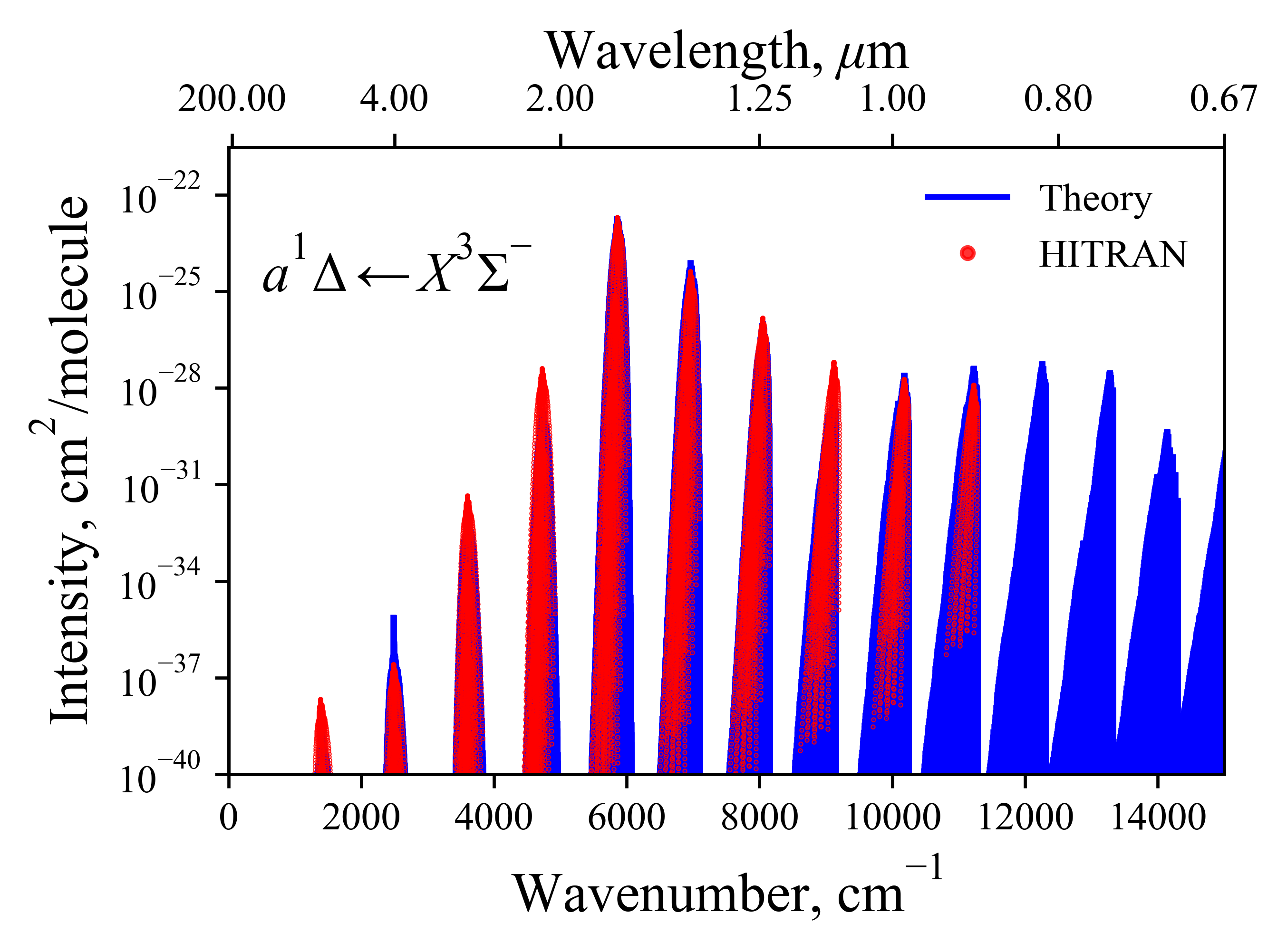}
    \caption{Comparison between the theoretical and HITRAN \band{\bstate}{\Xstate} (top) and \band{\astate}{\Xstate} (bottom) absorption spectrum for \wav{0-15000}. We simulate the spectra using a temperature of 296 K and scale the intensities by the fractional isotopologue abundance of 0.9479 \citep{jt857}.}
    \label{fig:bX_aX_HITRAN}
\end{figure}
\subsubsection{\hitran}
\label{subsubsec:HITRAN}
HITRAN produces empirical SO line lists which have been produced by fitting spectroscopic models to experimentally derived spectroscopic constants, lifetimes, and rotational branching ratios (ratios in perpendicular and parallel transition moments, see below discussion), hence we will compare to this data since it is the closest comparison of rovibronic intensities to experimental data for the \Xstate$-$\Xstate, \bstate$-$\Xstate, and \astate$-$\Xstate\ electronic bands.

\hitran\ \citep{jt857} provides empirical line list data on the first three electronic states of SO \Xstate, \astate, and \bstate\ to which we compare our theoretical spectra to. The HITRAN intensities for the \Xstate$\leftarrow$\Xstate\ band were originally presented by \citet{22BeJoLi.SO} and the forbidden $b\,^1\Sigma^+\leftarrow X\,{}^3\Sigma^-$ and $a\,^1\Delta\leftarrow X\,{}^3\Sigma^-$ bands are from \citet{21BeJoLi.SO}. In both studies, fitted spectroscopic constants from the literature were used to predict line positions, transition moments were obtained using LeRoy's LEVEL program \citep{LEVEL} which assumes the single state approximation, and their line lists were computed using PGOPHER \citep{PGOPHER}. \citet{21BeJoLi.SO} used $\Omega$-representation to allow for the single state approximation in line with LEVEL such that the forbidden band intensity are computed from effective dipoles between single $\Omega$-states, the so-called parallel and perpendicular transition moments, as opposed to the non-approximate intensity stealing mechanism via mixing of electronic wavefunctions through, e.g. SOCs, as we do. Perpendicular and parallel electronic transition moments between the spin-orbit states $b0^+-X0^+$, $b0^+-X1$, and $a2-X1$ (see appendix) were computed by \citet{21BeJoLi.SO} at an iC-MRCI/aug-cc-pCVQZ-DK level of theory and were scaled to the experimentally determined values by \citet{99SeFiRa.SO}. The HITRAN \Xstate$\leftarrow$\Xstate\ intensities were computed using the \ai\ ground state expectation dipole moment computed by \citet{22BeJoLi.SO} at a ic-MRCI+Q/ACV6Z-X2C/ED+Q level of theory.

In all comparisons below, we scale our computed intensities with the $\rm ^{32}S^{16}O$ isotopologue abundance 0.947926 given by \hitran\ \citep{jt857}. Figure \ref{fig:X_HITRAN} presents a comparison between our semi-empirical $X\,^3\Sigma^-\leftarrow X\,{}^3\Sigma^-$ rovibronic spectrum, where we compute stick spectrum using a temperature of 296~K, and the empirical \hitran\ line list \citep{22BeJoLi.SO}. We see good agreement in both the line positions and band structure where band intensities agree up to the fifth hot band at $\sim$ \wav{4500} where we compute higher intensities relative to the \hitran\ data. The $0\leftarrow0$ band agrees extremely well which can be seen in the sub plot of Fig.~\ref{fig:X_HITRAN}. The agreement in intensities confirms our methodology since the \hitran\ dipole was also scaled to the same experimental values discussed in Section \ref{subsec:int_scale}. The discrepancy  in intensities towards hotter bands can be attributed to the difference in the DMCs as well as the wavefunctions used to calculate the transition probabilities. 

% \begin{figure}
%     \centering
%      \includegraphics[width=\linewidth]{images/HITRAN_XX_0-7000_cm-1_comparison_bernath_X-dipole.png}
%     \caption{Comparison between the theoretical and HITRAN \band{\Xstate}{\Xstate} rovibrational band for  \wav{0-7000}. We simulate the spectra using a temperature of 296 K and scale the intensities by the fractional isotopologue abundance of 0.9479 \citep{jt857}.}
%     \label{fig:X_HITRAN}
% \end{figure}

Figure \ref{fig:bX_aX_HITRAN} compares our computed $b\,^1\Sigma^+\leftarrow X\,{}^3\Sigma^-$ spectrum simulated at a temperature of 296 K to the empirical \hitran\ \citep{jt857} $b\,^1\Sigma^+\leftarrow X\,{}^3\Sigma^-$ line list \citep{21BeJoLi.SO}. For this comparison, we filtered out the magnetic dipole transitions present in the \hitran\ line list, since we only calculate electric dipole transitions. The selection rules for magnetic dipole branches are the same except they follow the non-parity changing rule. We see that our model supplements the \hitran\ line list at both the higher and lower wavenumber regions ($<$ \wav{4000} and $>$ \wav{12500}) where line positions, band structure and intensities generally show good agreement which is to be expected since the HITRAN $b\,^1\Sigma^+\leftarrow X\,{}^3\Sigma^-$ dipoles were scaled to the same experimental values for the \bstate\ lifetime as we do. For vibrational bands energetically below $\sim$ \wav{7500} the two spectra begin to deviate from each other, where \LineList\ tends to be slightly lower than the \hitran\ intensities. However, this is below the standard \hitran\ intensity threshold of $10^{-30}$ cm$^2$/molecule, and so these bands are typically of less importance. For intensities above the threshold, we see good agreement in line with the methodologies used. 

We also compare to the forbidden $a\,^1\Delta\leftarrow X\,{}^3\Sigma^-$ band in Fig.~\ref{fig:bX_aX_HITRAN} to the theoretical \hitran\ line list at 296 K. The transitions here are all electric dipole in nature. We see the electronic band structure agrees well between the peak band up to the penultimate hot band before the \hitran\ terminus at $\sim$\wav{11000} with the largest intensity deviation being between vibronic bands outside the \wav{5000--8500} spectral range. The low energy bands $<$\wav{4500} all sit below the $10^{-30}$ cm$^2$/molecule intensity threshold, which are less important spectroscopically. Differences in the band intensity are difficult to disentangle since $a\,^1\Delta\leftarrow X\,{}^3\Sigma^-$ is a dipole forbidden band where intensities are accumulated through 'intensity stealing' mechanism via multiple coupling channels in our model.

The general agreement with the empirical \hitran\ line list confirms we are not missing any fundamental physics since we both produce similar spectra using a different methodology. Furthermore, \citet{21BeJoLi.SO} scale their $b\,^1\Sigma^+\leftarrow X\,{}^3\Sigma^-$ effective dipoles using the same lifetime provided by \citet{99SeFiRa.SO} as we did, whereas the $a\,^1\Delta\leftarrow X\,{}^3\Sigma^-$ band have no reliable experimental dipoles or lifetimes to scale the \ai\ dipoles to except the indirect \astate -- \astate\ dipole moment presented by \citep{70Saito.SO} which only we scale to. Agreement for the latter band then confirm the good quality of the PECs, SOCs, and (T)DMCs used.

% \begin{figure}
%     \centering
%     \includegraphics[width=\linewidth]{images/HITRAN_bX_comaprison.png}
%      \includegraphics[width=\linewidth]{images/HITRAN_aX_comaprison.png}
%     \caption{Comparison between the theoretical and HITRAN \band{\bstate}{\Xstate} (top) and \band{\astate}{\Xstate} (bottom) absorption spectrum for \wav{0-15000}. We simulate the spectra using a temperature of 296 K and scale the intensities by the fractional isotopologue abundance of 0.9479 \citep{jt857}.}
%     \label{fig:bX_aX_HITRAN}
% \end{figure}

\begin{figure*}
    \centering
    \includegraphics[width=\linewidth]{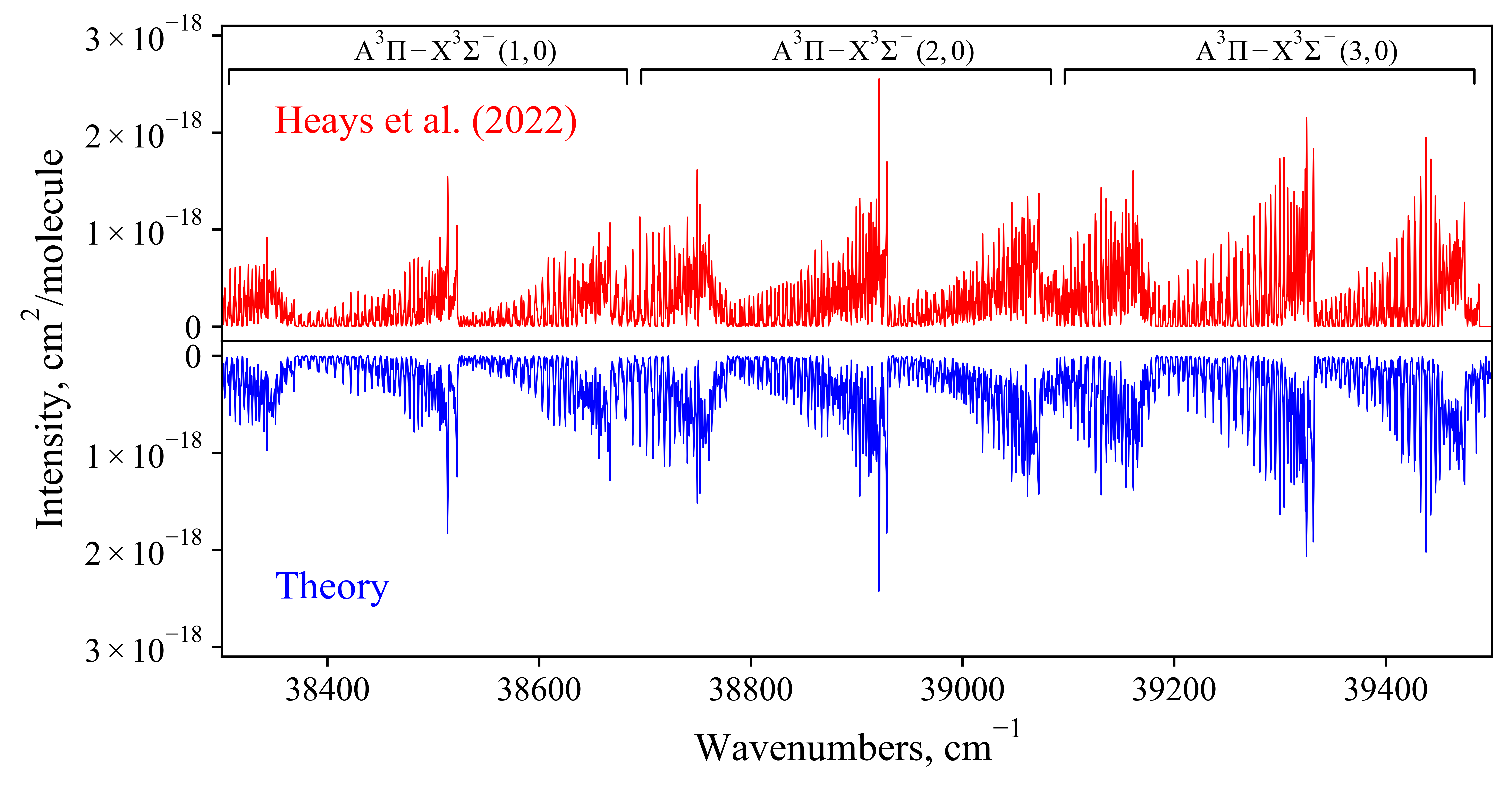}
    \caption{Comparison of our computed absorption \Astate$(v^{\prime}=1,2,3)-$\Xstate$(v^{\prime\prime}=0)$ band to the cross-sections generated from \citet{22HeStLy.SO} empirical line list fitted band-by-band to their measured spectra.}
    \label{fig:A-X_allBands_comparison}
\end{figure*}

% \begin{figure}
%     \centering
%     \includegraphics[width=\linewidth]{images/22HeStLy_comparison_A_X_1-0.png}
%      \includegraphics[width=\linewidth]{images/22HeStLy_comparison_A_X_2-0.png}
%      \includegraphics[width=\linewidth]{images/22HeStLy_comparison_A_X_3-0.png}
%     \caption{Comparison of the separated vibrational bands (top:$(1,0)$, middle:$(2,0)$, bottom:$(3,0)$) of our computed \Astate$-$\Xstate band to the cross sections generated from the band-by-band fitted line list of \citet{22HeStLy.SO} to their high resolution Fourier transform spectrum.}
%     \label{fig:A-X_vibBands_comparison}
% \end{figure}

\subsubsection{A--X  Bands}
\label{subsec:22HeStLy Comparison}
The recent study by \citet{22HeStLy.SO} measured, in high resolution, the FUV \Astate$-$\Xstate\ band via Fourier-transform spectroscopy up to the $(v',v")= (3,0)$ band for $J\leq51$; the \Bstate$-$\Xstate\ and \Cstate$-$\Xstate\ bands were also measured but we do not compare to these here. Heays {\it et al.} present an empirical line list where effective Hamiltonian spectroscopic constants were fitted band-by-band to their measured spectrum, providing quantum number assignments and oscillator strengths for each assigned transition. Their coupled-band models reproduce the experimentally measured line positions, intensities, and widths to within $5\%$ uncertainty. With this, we converted their line list to the \exocross\ format to compute corresponding cross-sections, which we compare to. For all spectra simulations a temperature of T = $360$ K and Gaussian line broadening of HWHM \wav{0.3} was used. Figure \ref{fig:A-X_allBands_comparison} shows the comparison between our computed \Astate$(v^{\prime}=1,2,3)-$\Xstate$(v^{\prime\prime}=0)$\ band intensities in blue (bottom panel) and the simulated band intensities of \citet{22HeStLy.SO} in red (top panel). We see excellent agreement in line positions, intensities, and band structure, where a mirror plot was chosen since overlaying the spectra made it hard to distinguish between them since they agree so well. 

% Spectra overlays of the individual vibrational bands can be seen in Figure \ref{fig:A-X_vibBands_comparison}, where indeed we see excellent agreement between our computed spectrum and experiment. 

We are confident that our model correctly reproduces the experimental spectra for the \Astate$(v^{\prime}=1,2,3)-$\Xstate$(v^{\prime\prime}=0)$\ band, confirming the good quality of our PECs, (T)DMCs, and couplings to other states.

\section{Conclusion}

We present the semi-empirical \LineList\ line list for $^{32}$S$^{16}$O constructed starting from the refinement of the \ai\ spectroscopic model presented by \citet{22BrYuKi.SO} to empirically derived energy levels, or \marvel\ energy levels. As part of the line list creation, a \marvel\ analysis of 29 experimental transition sources resulted in a self-consistent set of 8558 rotation-vibration energy levels ($J \leq 69$ and $v \leq 3$) for the \Xstate, \astate, \bstate, \Astate, \Bstate, \Cstate\ electronic states, where 48~972/50~106 experimental transitions were validated. 
The \LineList\ SO line list supplements existing ExoMol line lists for SO$_2$ \citep{jt635} and SO$_3$ \citep{jt641}.

The \Xstate\ state expectation value of the dipole moment operator was fitted to an analytical form and shown to improve the non-physical flattening of the vibrational transition moment and NIDL compared to using the grid interpolated form of the DMC. This resulted the physical exponential decay of the \band{\Xstate}{\Xstate} IR spectral band.

Comparison of the simulated rovibronic spectrum of SO to experiment/\hitran\ show good agreements in both positions and intensities. However, inspection of the $v=0\rightarrow0$ of the forbidden \band{\bstate}{\Xstate} band revealed disagreeing P- and R- branch ratios to experiment, where tuning of the spectroscopic model showed no proper inversion of the branch intensities. Analysis of the electronic wavefunctions revealed that the band intensities are dominated by competing diagonal $X$ and $b$ DMCs, which contribute to the parallel transition moment. The weaker perpendicular transition moment was shown to produce intensities of the desired P- and R- branch ratio, but was much weaker than the parallel transition component to the intensities. Analysis on the basis set revealed the branch ratio to be more sensitive when including a larger vibrational basis, but still does not produce the desired branch ratio.

The future work includes extension to the UV region with the \Bstate$\leftarrow{}$\Xstate and \Cstate$\leftarrow{}$\Xstate  electronic bands,
and production of photodissociation cross sections and rates.

\section*{Acknowledgements}

 This work was supported by the European Research Council (ERC) under the European Union’s Horizon 2020 research and innovation programme through Advance Grant number 883830 and  the STFC Projects No. ST/M001334/1 and ST/R000476/.

\section*{Data Availability}

The states, transition, opacity and partition function files for the SO line lists are provided via \href{https://exomol.com}{www.exomol.com}. The open access programs \duo\ and \textsc{ExoCross} are available from \href{https://github.com/exomol}{github.com/exomol}.

\section*{Supporting Information}

Supplementary data are available at MNRAS online, which include (a) the final spectroscopic model of SO is provided in the form of the \duo\ input file, containing all the curves, parameters as well as the MARVEL energy term values of SO used in the fit; (b) the MARVEL data set: the experimental transition frequencies and MARVEL energies.



\bibliographystyle{mnras}
\bibliography{./bib/journals_astro,./bib/SO,./bib/CO,./bib/atomic,./bib/exoplanets,./bib/ISM,./bib/linelists,./bib/methods,./bib/programs,./bib/MARVEL,./bib/SO2,./bib/stars,./bib/planets,./bib/abinitio,./bib/sy,./bib/CS,./bib/diabatisation,./bib/atmos,./bib/partition,./bib/Books,./bib/jtj,./bib/diatomic}

\begin{table}
\footnotesize
    \centering
    \caption{The largest expansion coefficients for the \Xstate\ and \bstate\ wavefunctions in the $\Omega$ representation. The column headers with a subscript '\duo' are the computed \duo-states which have components in the basis states given as rows due to spin-orbit coupling.}
    \label{tab:e_wavefunc_expansion}
    \begin{tabular}{lrrrr}
        \hline
        \multicolumn{1}{l}{Basis} & \multicolumn{1}{l}{$\ket{\Xstate_{0+}}_\duo$} & \multicolumn{1}{l}{$\ket{\bstate_{0}+}_\duo$} & \multicolumn{1}{l}{$\ket{\Xstate_{+1}}_\duo$} & \multicolumn{1}{l}{$\ket{\Xstate_{1-}}_\duo$}\\
        \hline\hline
$\ket{\Xstate_{-1}}$ & 0 & 0 & 0.99999 & 0 \\ 
$\ket{\Xstate_{+1}}$ & 0 & 0 & 0 & 0.99999 \\ 
$\ket{\Astate_{+1}}$ & 0 & 0 & 0 & -0.00119 \\ 
$\ket{\Astate_{-1}}$ & 0 & 0 & 0.00119 & 0 \\ 
$\ket{\Cstate_{+1}}$ & 0 & 0 & 0 & 0.00035 \\ 
$\ket{\Cstate_{-1}}$ & 0 & 0 & -0.00035 & 0 \\ 
$\ket{\Xstate_{0+}}$ & -0.99964 & -0.02670 & 0 & 0 \\ 
$\ket{\bstate_{0+}}$ & 0.02669 & -0.99964 & 0 & 0 \\ 
$\ket{\Astate_{0+}}$ & 0.00121 & -0.00093 & 0 & 0 \\ 
$\ket{\Astate_{0-}}$ & -0.00121 & 0.00093 & 0 & 0 \\ 
$\ket{\Cstate_{0+}}$ & -0.00034 & -0.00041 & 0 & 0 \\ 
$\ket{\Cstate_{0-}}$ & 0.00000 & 0.00040 & 0 & 0 \\ 
        \hline\hline
    \end{tabular}
\end{table}
\section*{Appendix: The \band{\bstate}{\Xstate} Band, an example of intensity contributions from a forbidden band}
%\label{subsubsec:intensity Steal}

\begin{figure}
    \centering
    \includegraphics[width=\linewidth]{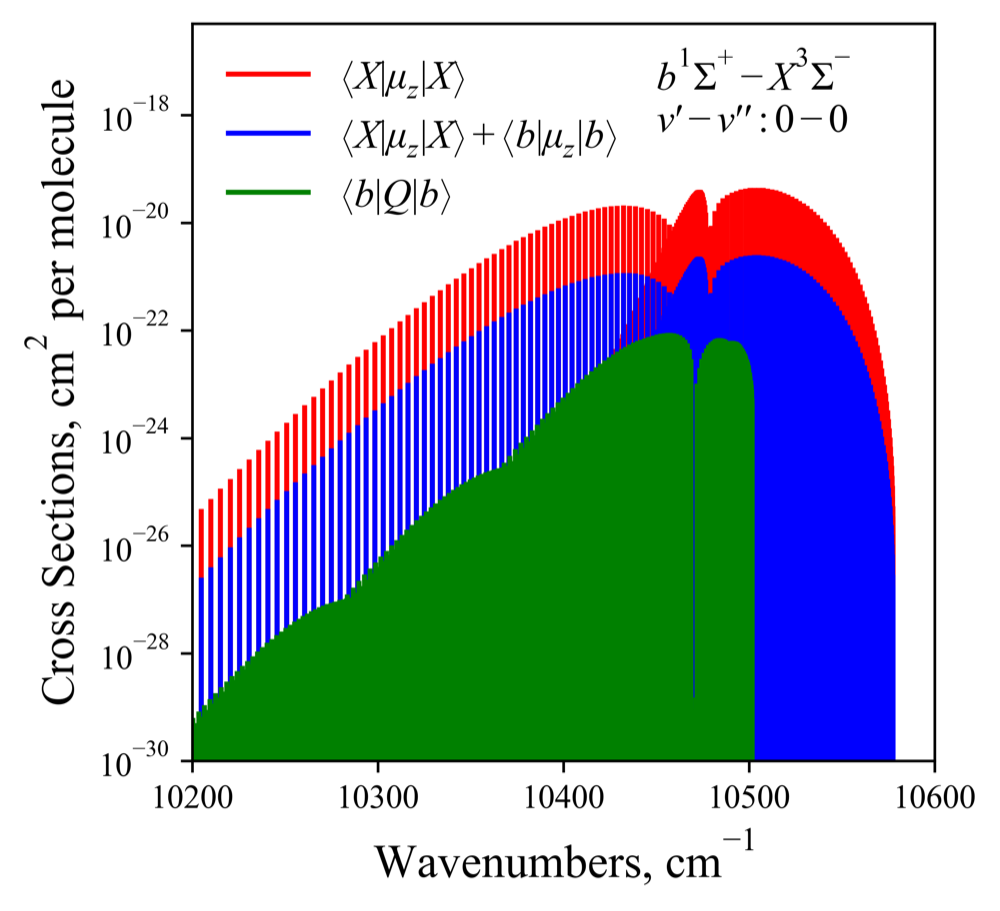}
    \caption{Visualisation of the different contributions to the forbidden \band{\Xstate}{\bstate} band intensities. 'Q' refers to the magnetic dipole moment, where the green spectra are due to magnetic dipole transitions, which is orders of magnitude weaker than the intensity stealing mechanism.}
    \label{fig:bX_spec}
\end{figure}

Dipole forbidden transitions can arise through multiple mechanisms, such as through the magnetic dipole moment (MDM), quadrupole moment, and from intensity stealing. Intensity stealing propagates through the mixture of electronic state wave-functions via couplings such as SOCs and EAMCs, where contributions to the forbidden intensities are derived through taking dipole matrix elements in the eigenstates of the diagonalised Hamiltonian constructed from the coupled $\Lambda-\Sigma$ basis. To understand these intensity contributions to the \band{\Xstate}{\bstate} band we performed an analysis of the  \duo\ computed electronic state wavefunctions corresponding to the eigensolutions of the diagonalised Hamiltonian which included SOCs, EAMCs, and DMCs for the full 11 state system described in \citet{22BrYuKi.SO} plus additional \brkt{\bstate}{{\rm SO}_x}{\Astate}, \brkt{\astate}{{\rm SO}_x}{\Astate} and \brkt{\bstate}{{\rm SO}_x}{\Cstate} couplings. The contributions to the \Xstate\ and \bstate\ computed wavefunctions in the $\Omega$ representation are shown in Table \ref{tab:e_wavefunc_expansion} which gives the expansion coefficients $C_n$ of the wavefunctions in the eigenbasis of the diagonalised Hamiltonian
\begin{equation}
    \Psi^{J,\tau}_{\Omega} = \sum_n C_n^{J,\tau}\ket{n}
\end{equation}
where $J$ is the rotational quantum number, $\tau$ is the parity, and $n$ represents the full set of quantum numbers $\ket{n}=\ket{{\rm State},J,\Omega,\Lambda,S,\Sigma,v}$. We see that because of large SOC between \Xstate\ and \bstate, they share sizeable contributions in their final mixed wavefunctions of their corresponding unmixed basis states. The amount of intensity stealing will then distribute itself through subsequent coupling of the dipole operators in the new mixed state basis. To this end, we consider the parallel and perpendicular transition dipole moments (TDMs) which couple $\Omega=0^{+}-0$ and $\Omega=0^{+}-1$ states, respectively. In spherical tensor form they read, considering the \Xstate\ and \bstate\ states,
\begin{equation} 
    \mu_0 = \bra{b^{1}\Sigma^{+}_0}\mu_z\ket{X^3\Sigma^{-}_0}
\label{eq:mu0}
\end{equation}
\begin{equation}
    \mu_{1} = \pm\bra{b^{1}\Sigma^{+}_0}2^{-1/2}(\mu_x\mp i\mu_y)\ket{X^3\Sigma^{-}_{\pm 1}}
\label{eq:mu1}
\end{equation}
The experimental measurement and analysis by \citet{99SeFiRa.SO} of the \band{\bstate}{\Xstate} emission band and work by \citep{21BeJoLi.SO} shows the $\mu_0$ and $\mu_{\pm 1}$ TDMs to be of the same order of magnitude. However, comparing the \band{\bstate}{\Xstate} $v=0-0$ band measured by \citet{99SeFiRa.SO} to our semi-empirical line list and the \hitran\ empirical line list show disagreements in the P- and R- branch ratios where experiment predicts the P- branch have the most intensity within the band. Considering the intensity stealing mechanism, cross-examining Eqs.(\ref{eq:mu0},\ref{eq:mu1}) with Table \ref{tab:e_wavefunc_expansion} reveals that intensity stealing is different for the perpendicular and parallel transitions and therefore the branch ratios for these transitions. This could explain the experimental observation in \citet{99SeFiRa.SO} that the P- and R- branch ratios are different. In an attempt to understand this discrepancy, our analysis on the mixed \Xstate\ and \bstate\ state wavefunctions reveals that the competition between the \brkt{\bstate}{\mu_z}{\bstate} and \brkt{\Xstate}{\mu_z}{\Xstate} DMCs provides the dominant contributions to $\mu_0$, as facilitated through the large \brkt{\bstate}{{\rm SO}_z}{\Xstate} spin-orbit coupling, where \brkt{\bstate}{\mu_z}{\bstate} subtracts from \brkt{\Xstate}{\mu_z}{\Xstate} and can be seen in Figure \ref{fig:bX_spec} as the reduction of intensity between the red and blue spectra. $\mu_1$, however, has leading contributions from the \brkt{\Astate}{\mu_x}{\Xstate} dipole as facilitated through the \brkt{\bstate}{{\rm SO}_x}{\Astate} and \brkt{\Xstate}{{\rm SO}_x}{\Astate} couplings. We find that the perpendicular TDM $\mu_1$ is responsible for the P- R- branch asymmetry ($P>R$) as seen in the Fourier transform spectroscopy by \citet{99SeFiRa.SO} whereas the parallel TDM $\mu_0$ produces the opposite branch ratios ($P<R$). However, when considering only a vibrationless expansion for our wavefunctions, that is $v = 0$, the $\mu_1$ TDM is very weak producing much lower intensities than the parallel transition moment and as a result the \band{\bstate}{\Xstate} band has a $P<R$ branch ratio. When considering a larger vibrational basis ($v=20$) the P- R- branch ratio becomes more sensitive to changes in DMCs and SOCs. From this analysis we see that the vibrational TDMs have an effect on the rovibrational band intensity distribution in favour of the P- R- branch ratio as seen in \citet{99SeFiRa.SO}.

\bsp	% typesetting comment
\label{lastpage}
\end{document}